\documentclass{aa}  

\usepackage{graphicx}
\usepackage{txfonts}
\usepackage[]{hyperref}
\usepackage{natbib}
\bibpunct{(}{)}{;}{a}{}{,} 
\usepackage{amsmath}    
\usepackage{bm}
\usepackage{xcolor}
\usepackage[shortlabels]{enumitem}
\usepackage{float}
\usepackage{placeins}

\begin{document}

\title{Collisional damping in debris discs: Only significant if collision velocities are low}
\titlerunning{Collisional damping in debris discs}

\subtitle{}

\author{Marija R. Jankovic\inst{1},
Mark C. Wyatt\inst{2}
\and
Torsten L\"{o}hne\inst{3}
}

\institute{Institute of Physics Belgrade, University of Belgrade, Pregrevica 118, 11080 Belgrade, Serbia\\
\email{marija.jankovic@ipb.ac.rs}
\and
Institute of Astronomy, University of Cambridge, Madingley Road, Cambridge CB3 0HA, UK
\and
Astrophysikalisches Institut und Universit\"{a}tssternwarte, Friedrich-Schiller-Universit\"{a}t Jena, Schillerg\"{a}ßchen 2-3, D-07745 Jena, Germany
}

\date{}

 
\abstract
{Dusty debris discs around main sequence stars are observed to vary widely in terms of their vertical thickness. Their vertical structure may be affected by damping in inelastic collisions. Although kinetic models have often been used to study the collisional evolution of debris discs, these models have not yet been used to study the evolution of their vertical structure.}
{We extend an existing implementation of a kinetic model of collisional evolution to include the evolution of orbital inclinations and we use this model to study the effects of collisional damping in pre-stirred discs.}
{We evolved the number of particles of different masses, eccentricities, and inclinations using the kinetic model and used Monte Carlo simulations to calculate collision rates between particles in the disc. We considered all relevant collisional outcomes including fragmentation, cratering, and growth.}
{Collisional damping is inefficient if particles can be destroyed by projectiles that are of much lower mass. If that is the case, catastrophic disruptions shape the distributions of eccentricities and inclinations, and their average values evolve slowly and at the same rate for all particle sizes.}
{The critical projectile-to-target mass ratio ($Y_{\rm c}$) and the collisional timescale jointly determine the level of collisional damping in debris discs. If $Y_{\rm c}$ is much smaller than unity, a debris disc retains the inclination distribution that it is born with for much longer than the collisional timescale of the largest bodies in the disc. Such a disc should exhibit a vertical thickness that is independent of wavelength even in the absence of other physical processes. Collisional damping is efficient if $Y_{\rm c}$ is of order unity or larger. For millimetre-sized dust grains and common material strength assumptions, this requires collision velocities of lower than $\sim$\,40\,m\,s$^{-1}$. We discuss the implications of our findings for exo-Kuiper belts, discs around white dwarfs, and planetary rings.}


\maketitle



\section{Introduction}
Dusty discs have been detected around numerous main sequence stars. These are believed to be merely the observable component of discs populated with larger bodies, planetesimals, and perhaps even dwarf planets. Such systems are commonly called debris discs \citep{Wyatt2008,Hughes2018}. In debris discs, planetesimals and dust are respectively at the top and bottom of a collisional cascade, a system in which collisions and fragmentation of larger objects resupply the abundance of smaller objects. Thus, through characterisation of the small dust in these discs, it is possible to probe the structure of the large planetesimals. The vertical structure of these discs in particular can be useful to constrain their dynamical state, as the vertical profile of a disc reveals the inclination distribution of the orbits of its particles, and this distribution is directly related to the velocities at which particles collide. That is, the vertical structure and the dynamical state of a debris disc are closely related to its collisional evolution and the transfer of mass down the collisional cascade.

Observationally, the vertical structure of debris discs is most easily probed for discs that are seen near edge-on, for which a disc scale-height or aspect ratio can be measured. Measurements so far show a great diversity among discs. At millimetre (mm) wavelengths, aspect ratios range from 0.02 for AU Mic \citep{Terrill2023} and 0.08 for HR 4796 \citep{Kennedy2018}, up to 0.13 for HD 16743 \citep{Marshall2023} and 0.13-0.28 for HD 110058 \citep{Hales2022}. At wavelengths in the near-infrared (NIR), for which the highest resolutions have been achieved, it has been shown that discs are diverse in terms of  not only their aspect ratios
, but also the shape of their vertical density profile \citep{Olofsson2022}. Vertical profile shapes at mm wavelengths can also be nontrivial, as found for the disc around $\beta$ Pic \citep{Matra2019}. In some systems, the aspect ratio may also vary with distance from the star, which can be inferred with sophisticated nonparametric methods \citep{Han2022}.

Discs observed at multiple wavelengths can provide a substantial amount of information, as they probe the scale-height of particles of different sizes in a single dynamical system. So far, there appears to be diversity in these results as well. For example, for HR 4796 (HD109573), the aspect ratio inferred from ALMA observations of 0.08 \citep[at the wavelength of 880\,$\mu$m;][]{Kennedy2018} is comparable to  the SPHERE value of 0.06 \citep[at the wavelength of $\sim$\,0.7\,$\mu$m;][]{Olofsson2022}. On the other hand, in AU Mic the aspect ratio appears to be increasing as a function of wavelength \citep{Olofsson2022,Daley2019,Vizgan2022}. All in all, the amount of available observational data on the vertical structure of debris discs is steadily growing, and theoretical models are needed to understand them.

It has long been argued that the velocity evolution of  a debris disc (and thus the evolution of its scale-height) is governed by the largest bodies in the disc (the largest planetesimals or perhaps even dwarf planets), which increase the inclinations and eccentricities of all the bodies through gravitational interactions, a process known as viscous stirring \citep{Kenyon2002b, Kenyon2004a, Kenyon2008, Kenyon2010, Kennedy2010, Krivov2018b}. In the absence of damping mechanisms (or if these are inefficient), an initially thin disc will be viscously stirred by the largest bodies in the disc, so that its scale-height (and the particles' inclinations, eccentricities, and the velocity dispersion) may depend on the age of the system, as well as on the properties of the bodies doing the stirring. Thus, assuming the absence of damping mechanisms, the properties of the largest bodies in the disc can be constrained through measurement of the scale-height \citep{Nakano1988, Artymowicz1997, Quillen2007, Matra2019}.

One damping mechanism is certain to exist: collisions among particles are inelastic and this has the effect of damping particle inclinations and eccentricities over time. The efficiency with which collisional damping impedes viscous stirring depends on the size distribution of the bodies in the belt \citep{Lissauer1993}. Using an analytic model, \citet{Pan2012} proposed that there could be an equilibrium between viscous stirring and collisional damping in a collisional cascade, such that, among small dust grains, particle velocity increases with particle size. On the other hand, time-dependent simulations of debris-disc evolution that account for both viscous stirring and collisional damping, as well as continued accretion of the largest bodies \citep{Kenyon2002b, Kenyon2004a, Kenyon2008, Kenyon2010}, typically show collisional damping to only be important early on, while viscous stirring dominates at later stages due to efficient (dwarf) planet formation, making the velocity independent of particle size for the small dust. However, the results vary based on the exact mixture of planetesimals and pebbles that the disc initially consists of \citep{Najita2022}, meaning that in these simulations there can be a phase in the disc evolution during which particle velocity increases as a function of particle size. Therefore, there is no clear consensus on the degree to which collisional damping may affect the debris-disc scale-heights that are being observed.

As physical collisions are one of the most fundamental processes in debris discs, it is important to understand their role in the evolution of particle eccentricities and inclinations (and thus in the evolution of the disc scale-heights). One numerical method in particular is well suited for this --the kinetic approach \citep{Krivov2005}, whereby the collisional and dynamical evolution of particles are considered in a fully coupled manner in the parameter space of particle size and selected orbital elements. This method allows arbitrary distributions of orbital elements to be considered on a discretised grid. In comparison, in the mixed statistical approach \citep[e.g.][]{Kenyon2002a,Kenyon2002b,Kenyon2004a}, the evolution of mass and the evolution of velocity are considered concurrently but separately \footnote{For the most part, some redistribution of the kinetic energy of the colliders to their fragments can be included in this method directly within the collisional outcome prescription \citep{Kenyon1999}.}, and it is assumed that the functional form of the distributions of eccentricities and inclinations is fixed. However, the kinetic approach has so far not been used to study the evolution of particle inclinations or debris disc scale-heights.

In this paper, we aim to study the effects of physical collisions on particle eccentricities and inclinations in debris discs  in detail. To do so, we extend a previous implementation of the kinetic approach \citep{vanLieshout2014} to include particle inclinations in the parameter space, using Monte Carlo simulations \citep{Wyatt2010} to calculate particle collision rates. There are two major aims of this paper: to present this extended numerical model and to study the importance of collisional damping in collisional cascades. We start by discussing the underlying physics of collisional damping in Sect. \ref{sec:theory}, and present the methods of our numerical approach in Sect. \ref{sec:methods}. In Sects. \ref{sec:results} and \ref{sec:discussion} we present and discuss our results. Finally, in Sect. \ref{sec:conclusions}, we summarise our findings.

\section{Theoretical expectations} \label{sec:theory}
In this section we discuss some of the well-known results from the literature about collisional damping and we discuss collisional damping in collisional cascades whose evolution is governed by destructive collisions. We are only concerned with physical collisions and not with close encounters and gravitational scattering. A simple model of such a physical collision is a collision of two indestructible spheres, with the normal and the transversal components of the relative velocity at the point of impact changing in the collision proportionally to coefficients of restitution \citep[e.g.][]{Trulsen1971,Brahic1977,BrahicHenon1977,Hornung1985}. We further simplify this model by assuming physical collisions are completely inelastic.
The total kinetic energy of colliders pre-collision is a sum of the energy associated with centre-of-mass motion and with their relative velocity.
In a completely inelastic collision the latter term is entirely dissipated and post-collision both bodies are moving at their centre-of-mass velocity. This assumption may not be fully realistic for all types of collisional outcomes in real discs, but this does not affect the main takeaways of this work (see the discussion in Sect. \ref{sec:limitations}). Thus, we leave exploration of more detailed prescriptions for future work.

Even a completely inelastic collision does not always lead to a lower velocity for both bodies. For example, if a population of larger bodies has higher velocity dispersion than a population of smaller bodies, it will collisionally stir the smaller bodies. But, each collision loses energy and the overall effect is collisional damping, meaning, reduction in the velocity dispersion. In a circumstellar disc in which the velocity dispersion arises from particles having different eccentricities and inclinations, this means a decrease in the typical particle orbital eccentricity and inclination with time \citep{Brahic1977,Hornung1985}. 

We wish to estimate the rate at which the velocity dispersion and the typical eccentricity and inclination decrease with time. To do so, we first consider the rate at which particles collide in a circumstellar disc, which can be easily estimated with a particle-in-a-box model \citep[e.g.][]{Wyatt2007a,Rigley2020}. In such a model we can consider a power-law size distribution of particles, $n(s) \textrm{d}s\propto s^{-q} \textrm{d}s$, with a bulk material density $\rho_{\rm b}$, a total mass $M$, and with minimum and maximum radius of $s_{\rm min}$ and $s_{\rm max}$, respectively. This material is assumed to be spread out over a circumstellar belt centred at stellocentric radius $a$ with a width $\Delta a$ and average inclination $\langle i \rangle$. The volume of the disc can then be estimated as $V \sim 4 \pi a^2 \Delta a \langle i \rangle$. The average impact velocity depends, in general, on the average eccentricity and the average inclination of particle orbits \citep[e.g.][]{Lissauer1993}. Assuming the ratio between the average eccentricity and the average inclination to be a constant of order unity, the average impact velocity can be estimated as $v_{\rm rel} \sim v_{\rm kep}(a) \langle i \rangle$. The rate at which a target particle of size $s_{\rm t}$ collides with other particles in the size range of $s_{\rm min}$ to $s_{\rm max}$ is then given by
\begin{equation} \label{eq:r_coll}
    R_{\rm coll}(s_{\rm t}) = C \int_{s_{\rm min}}^{s_{\rm max}}  s_{\rm p}^{-q} (s_{\rm p} + s_{\rm t})^2 \textrm{d}s_{\rm p},
\end{equation}
where subscript p stands for the projectile particle colliding with the target and
\begin{equation} \label{eq:r_coll_constant}
    C = \frac{3(4-q)}{16\pi} (G M_*)^{0.5} \rho_{\rm b}^{-1} a^{-3.5} \left( \frac{\Delta a}{a} \right)^{-1} M s_{\rm max}^{q-4}.
\end{equation}

Next, we account for the fact that different collisions contribute differently to collisional damping, because the degree of velocity damping depends on the mass ratio of the colliders. For a rigorous, more general derivation based on kinetic theory we refer the reader to \citet{Hornung1985}. Here we merely sketch out how this dependence on the mass ratio arises by deriving an order-of-magnitude estimate, assuming all particles to have an identical Gaussian velocity distribution. In a thin circumstellar disc, the velocity dispersion ($\langle \delta v^2 \rangle$) around the Keplerian velocity ($v_{\rm K}(a)$) arises from the eccentricity and inclination dispersions ($\langle e^2 \rangle$ and $\langle i^2 \rangle$). In this simplified model we also assume that the ratio of the eccentricity and the inclination dispersion is a constant. In a collision of a target of mass $m_{\rm t}$ with a projectile of mass $m_{\rm p}$, $\langle e^2 \rangle$ and $\langle i^2 \rangle$ change by a factor of the order of the fraction of the average kinetic energy associated with the random velocities that is lost on average in a single collision ($\langle \Delta \delta v_{\rm t}^2 \rangle / \langle \delta v_{\rm t}^2 \rangle$). We can calculate this fraction by considering a single collision in which particles collide with velocities $\mathbf{v_{\rm t}}$ and $\mathbf{v_{\rm p}}$ (in the inertial frame of reference), in which the energy loss for the target is given by 
\begin{align}
    \frac12 m_{\rm t} v_{\rm t}^2 - \frac12 m_{\rm t} v_{\rm CM}^2 = \nonumber \\ \frac12 \frac{m_{\rm t}}{(m_{\rm t}+m_{\rm p})^2} 
    \left( m_{\rm p}^2 (v_{\rm t}^2 - v_{\rm p}^2) + 2 m_{\rm t} m_{\rm p} \mathbf{v_{\rm t}} (\mathbf{v_{\rm t}} - \mathbf{v_{\rm p}}) \right). 
\end{align}
Averaging the above expression over the velocity distributions of the colliders, using our assumptions of Gaussian velocity distributions independent of particle size and disregarding a numerical factor of order unity, the fraction we are looking for is given by
\begin{equation}
    \frac{ \langle \Delta \delta v_{\rm t}^2 \rangle }{ \langle \delta v_{\rm t}^2 \rangle } \sim \frac{m_{\rm t}m_{\rm p}}{(m_{\rm t}+m_{\rm p})^2}.
\end{equation}
It follows that the eccentricity, the inclination and the velocity dispersion decrease over time approximately at a rate given by
\begin{equation} \label{eq:r_damp}
    R_{\rm damp}(s_{\rm t}) = C \int_{s_{\rm min}}^{s_{\rm max}}  s_{\rm p}^{-q} (s_{\rm p} + s_{\rm t})^2 \frac{m_{\rm t} m_{\rm p}}{(m_{\rm t} + m_{\rm p})^2} \textrm{d}s_{\rm p}.
\end{equation}

From this it can be shown that collisions with equal-size bodies dominate collisional damping. It is useful to consider the number of particles in a logarithmic size bin, that is, to change the variable of integration to $\textrm{ln} s_{\rm p}$. For $q<4$, one obtains an integrand that is a non-monotonous function peaking approximately at $s_{\rm p} \sim s_{\rm t}$. This is to be intuitively expected. A much smaller projectile cannot change significantly the velocity of a massive target, while, on the other hand, for $q < 4$ smaller particles dominate the cross-section area and collisions with more massive `projectiles' occur at a much lower rate.

If it is further assumed that particle velocities are \textit{only} damped in collisions with particles in the same logarithmic unit size bin \citep{Pan2012},
\begin{align} \label{eq:r_damp_approx}
    R_{\rm damp}(s_{\rm t}) &= C \int_{s_{\rm min}}^{s_{\rm max}}  s_{\rm p}^{1-q} (s_{\rm p} + s_{\rm t})^2 \frac{s_{\rm t}^3 s_{\rm p}^3}{(s_{\rm t}^3 + s_{\rm p}^3)^2} \textrm{d ln}s_{\rm p} \nonumber \\
    &\sim C s_{\rm t}^{3-q}.
\end{align}
For the standard value of $q=3.5$ \citep{Dohnanyi1969}, $R_{\rm damp}(s_{\rm t}) \propto s_{\rm t}^{-1/2}$. This simple size-dependence is valid as long as $s_{\rm t}$ is not too close to either $s_{\rm min}$ or $s_{\rm max}$. One clear consequence of this size-dependence is that the velocity distributions of particles of different sizes do not remain identical, and the damping rate itself will change over time. Nevertheless, eq. (\ref{eq:r_damp}) provides a simple estimate for how quickly the eccentricities and the inclinations should change initially in a disc in which particles of all sizes are pre-stirred to the same velocity dispersion.

The above damping rate is applicable if the particle remains intact after each collision. But, for the debris discs to be observable, there must be destructive collisions producing dust. The change in particle mass in collisions is thus critical. As with damping, only a fraction of all collisions are important. For the body to be fragmented, the impact energy has to be greater than some size- and velocity-dependent critical specific energy $Q_{\rm D}^*$ \citep[e.g.][]{Benz1999,Stewart2009}. Equivalently, for some typical impact velocity in the disc, a target can only be destroyed by a projectile larger than some critical size $s_{\rm c}$. For projectiles that are much less massive than the target the impact energy is roughly the kinetic energy of the projectile and
\begin{equation} \label{eq:crit_ratio}
    \frac{s_{\rm c}}{s_{\rm t}} = \left( \frac{2 Q_{\rm D}^*}{v_{\rm rel}^2} \right)^{\frac{1}{3}}.
\end{equation}
The rate of destructive collisions is then given by
\begin{equation} \label{eq:r_frag}
    R_{\rm frag}(s_{\rm t}) = C \int_{s_{\rm c}}^{s_{\rm max}}  s_{\rm p}^{-q} (s_{\rm p} + s_{\rm t})^2 \textrm{d}s_{\rm p}.
\end{equation}
For $q>1$, the fragmentation rate is dominated by collisions with bodies of critical size $s_{\rm c}$. Analogous to eq. (\ref{eq:r_damp_approx}), not too close to either the top or the bottom of the cascade, the fragmentation rate is approximately given by
\begin{equation} \label{eq:r_frag_approx}
    R_{\rm frag}(s_{\rm t}) \sim C s_{\rm c}^{1-q} (s_{\rm c} + s_{\rm t})^2.
\end{equation}

It follows directly that, at the very least, the damping rate from eq. (\ref{eq:r_damp}) should be modified so that the upper limit of the integral equals $s_{\rm c}$, if $s_{\rm c} < s_{\rm t}$. The rate $R_{\rm damp}$ is then dominated by bouncing or cratering collisions with bodies just below the critical projectile size. More importantly, the fragmentation rate is then higher than the damping rate. If $s_{\rm c} \lesssim s_{\rm t}$, particles are destroyed faster than their velocities are damped\footnote{In the limiting case of $s_{\rm t} = s_{\rm max}$ for which the fragmentation rate diminishes for $s_{\rm c}=s_{\rm t}$, $R_{\rm frag}>R_{\rm damp}$ for $s_{\rm c}$ below $\approx 0.6 s_{\rm t}$.}. Their velocities are then inherited from their parent bodies and change very little before mass proceeds down the collisional cascade. Collisional damping would occur jointly with the transfer of mass down the collisional cascade, although if $s_{\rm c} \ll s_{\rm t}$ fragments will simply inherit the velocity distribution of the more massive of two colliders. Additionally, the more the impact energy exceeds the critical energy, the smaller the fragments and the further down the cascade the target's velocity is copied, hence equalising the velocities across the cascade. All in all, we expect that collisional damping is inefficient if the critical projectile-target size ratio is much smaller than unity. Velocities may still evolve in this case, but their evolution would depend on how the destruction and production rates shape the distributions of eccentricities and inclinations.

Now, in some discs it may happen that $s_{\rm c} \gtrsim s_{\rm t}$. In that case $R_{\rm frag} \lesssim R_{\rm damp}$, and the average eccentricity and inclination of particles would indeed fall off at the rate $R_{\rm damp}$ (provided these collisions are fully dissipating as assumed here). The key question is then if collisions still produce a collisional cascade and observable dust. For example, eq. (\ref{eq:crit_ratio}) is in fact no longer applicable, since it is derived under the assumption that $s_{\rm c} \ll s_{\rm t}$. For large values of $2 Q_{\rm D}^*/v_{\rm rel}^2$, the impact energy is insufficient to disrupt the target mass for any projectile mass. On the other hand, in this simplified model we only consider collisions at the average/typical collision velocity $v_{\rm rel}$, whereas particles in real discs have a distribution of velocities. Moreover, cratering/erosion can still produce dust in this case. In fact, studies have found that even when fragmenting collisions are present, cratering collisions can dominate the mass flux through the cascade \citep{Thebault2003,Thebault2007,Kobayashi2010}. All in all, to investigate this further we need a more comprehensive model, which we present in the next section.

Lastly, collisional damping also affects particles' semi-major axes. As particles lose energy in inelastic collisions, their semi-major axes are expected to shrink over time. More precisely, most of the mass can be expected to shift inwards, while some of the particles will end up on wider orbits, to conserve total angular momentum. This evolution of semi-major axes happens on much longer timescales than the damping of eccentricity and inclination, as found in N-body simulations \citep{Brahic1977}. N-body simulations of \citet{Lithwick2007} also support this large difference in timescales, and they also find analytically that the radial diffusion timescale of the ring will be longer than the damping timescale by a factor of $(\Delta a / (a e))^2$, where $e$ is the average eccentricity. For wide enough rings, therefore, it is justified to neglect the evolution of semi-major axis while focusing on the short-term evolution of eccentricities and inclinations. In numerical simulations presented in our work, $\Delta a / (a e) \gg 1$ in all physical models where damping is efficient. To further support neglecting semi-major axis evolution, note that \citet{Pan2012} argued that the ratio of the timescale of inclination damping to the timescale of semi-major axis damping is equal to the fraction of the orbital energy lost in a single collision, which is roughly the average inclination of particle orbits. That is typically observed to be less than $\sim 0.1$ in debris discs \citep[e.g.][]{Terrill2023,Kennedy2018,Olofsson2022,Daley2019}. Based on these arguments, we neglect the evolution of semi-major axis in this work. However, we also note that in the N-body simulations of \citet{Brahic1977} and \citet{Lithwick2007} all bodies are of equal size. Damping rates are size-dependent and in a collisional cascade of bodies spanning many orders of magnitude in size it may indeed occur that semi-major axes of the smallest bodies evolve faster than the evolution of the eccentricities and inclinations of the largest bodies. We further discuss this possibility in Sect. \ref{sec:discussion}.

\section{Methods} \label{sec:methods}
To model collisional evolution of particles in a circumstellar disc, we employ a modified version of the kinetic model implemented by \citet{vanLieshout2014} \citep[based on the approach developed by ][]{Krivov2005,Krivov2006,Lohne2008}. The spatial and size distributions of particles ($n$) in a phase space of particle masses ($m$) and orbital elements ($\bm{k}$) are evolved using the continuity equation
\begin{equation} \label{eq:cont_eq}
    \frac{\textrm{d}n}{\textrm{d}t} = \left( \frac{\textrm{d}n}{\textrm{d}t} \right)_{\rm gain} - \left( \frac{\textrm{d}n}{\textrm{d}t} \right)_{\rm loss}
,\end{equation}
where the gain (source) and the loss (sink) terms describe changes in the distribution function due to collisions. 
In this study we focus on the evolution of large particles in the outer regions of gas-free debris discs, and so we do not consider P-R drag, gas drag, nor sublimation of dust. 
To solve the continuity equation, the phase space $(m, \textbf{k})$ is discretised into a grid of bins, and the distribution function is discretised as a number of particles in each bin. The continuity equation becomes a series of ODEs.

The key assumption of the kinetic approach is that the collisional timescales, as well as the timescales of the drag processes, are much longer than orbital timescales. We further assume that the disc is axisymmetric. Under these assumptions, the evolution of the dust can be followed in a reduced phase space consisting of 4 variables: particle mass ($m$), eccentricity ($e$), semi-major axis ($a$) (or, alternatively, periastron distance) and inclination ($i$). The remaining orbital elements, which determine the orientations of the orbits, can be averaged over in the treatment of particle collisions.

The inclusion of inclination in the kinetic model phase space is the key difference between the model used in this work and the model used by \citet{vanLieshout2014}. The latter assumed that the particle inclination distribution was fixed. This allowed to use a simplified, analytic approach to calculate collisional probabilities and other relevant quantities \citep{Krivov2006}. To consider collisions between particles with arbitrary inclinations, we perform Monte Carlo simulations which sample the distributions of particle velocities and collision locations \citep{Wyatt2010}. In this work, we also account for a greater variety of collisional outcomes, including erosion and growth, bringing the model up-to-date with state-of-the-art debris disc kinetic models in this respect \citep[e.g.][]{Lohne2017}. We also make minor corrections to the code with regards to the calculation of collision terms.

Our treatment of particle collisions is detailed in Sect. \ref{sec:methods_collisions}. While the model is implemented to work with a four-dimensional phase space, due to computational costs we limit ourselves to using only a single semi-major axis bin. In other words, semi-major axis does not evolve in the results presented here (but we still take into account the reduction in the radial width of a belt if particle eccentricities are damped). This should not affect the main findings of our study, which concern chiefly the evolution of particles' eccentricities and inclinations, because the evolution of semi-major axis due to collisional damping happens on much longer timescales \citep[][see discussion in Sect. \ref{sec:theory}]{Brahic1977}. The exact phase-space grid we use is defined in Sect. \ref{sec:grid}. Several tests of our model are considered in Appendices \ref{sec:tests_mc} and \ref{sec:tests_km}.

\subsection{Collisions} \label{sec:methods_collisions}
First, we discuss the collision terms in the kinetic model and introduce the quantities characterising the collisions. Then, we discuss the calculation of the collisional probabilities and impact velocities using the Monte Carlo method, and finally the collisional outcomes considered in this model. Figure \ref{fig:flowchart} shows the overview of how the kinetic model and the Monte Carlo simulations are coupled.

\subsubsection{Collision terms in the kinetic model}
Without the transport terms, in the discretised form of the continuity equation, the number of particles in each bin evolves according to
\begin{equation}
     \frac{\textrm{d}n_i}{\textrm{d}t} = \sum_p \sum_t G(p, t, i) - \sum_p L(p, i) ,
\end{equation}
where indices $i$, $p$ and $t$ numerate bins in the phase space, and $G$ and $L$ denote the gain and the loss rates, respectively. Essentially, the first term on the right-hand side loops over various combinations of targets ($t$) and projectiles ($p$) whose collisions produce fragments in the bin with index $i$, and the second term loops over various projectiles ($p$) which can destroy or crater particles in the bin with index $i$. A factor 1/2 is applied for $p=t$ collisions in the first term and for $p=i$ collisions in the second term. The gain ($G$) and the loss ($L$) rates are averaged over all of the orbital elements not included in our phase space. This averaging is performed using a Monte Carlo simulation, which samples the distributions of the redundant orbital elements, producing a large number of particle pairs colliding at different locations and velocities. The gain and loss rates are given by
\begin{align}
    L(p, t) = &n_p n_t \sigma(m_p, m_t) \sum_j \Delta(\bm{k}_p, \bm{k}_t; j) \nonumber \\
    &\times v_{\rm imp}(m_p, \bm{k}_p, m_t, \bm{k}_t; j) \nonumber \\
    &\times c(m_p, \bm{k}_p, m_t, \bm{k}_t; j), \label{eq:disc_loss} \\
    G(p, t, i) = &n_p n_t \sigma(m_p, m_t) \sum_j  \Delta(\bm{k}_p, \bm{k}_t; j) \nonumber \\
    &\times v_{\rm imp}(m_p, \bm{k}_p, m_t, \bm{k}_t; j) \nonumber \\ 
    &\times c(m_p, \bm{k}_p, m_t, \bm{k}_t; j) \nonumber \\
    &\times f(m_p, \bm{k}_p, m_t, \bm{k}_t, m_i, \bm{k}_i; j) , \label{eq:disc_gain}
\end{align}
where the sum over $j$ counts over particle pairs produced in a Monte Carlo simulation. Here, $\sigma$ is the collisional cross section, $\Delta$ is the geometric probability of the collision \citep[similar in its meaning to the $\Delta$-integral of][]{Krivov2006}, $v_{\rm imp}$ is the impact velocity, $c$ is the probability that the collision results in loss of mass and $f$ is the probability that a collision of a particular target and projectile will result in the formation of a fragment with phase-space variables $m_i$ and $\bm{k}_i$ (see Sect. \ref{sec:methods_outcomes}). The probability $\Delta$ and the impact velocity are obtained from the Monte Carlo simulation, as described in the following section. The impact velocity only depends on particle masses for small dust particles affected by radiation pressure; in the results presented in this paper that is neglected.  

\subsubsection{Collisional probabilities and impact velocities} \label{sec:methods_mc}
To calculate geometric collision probabilities and particle velocities, and ultimately the gain and the loss rates due to collisions, we use a Monte Carlo approach based on the work of \citet{Wyatt2010}. The goal is to characterise collisions between two populations of particles with given semi-major axes (or pericentre distances), eccentricities and inclinations. With these orbital elements fixed, particles still collide at a range of distances from the central star (i.e. at different radii) and at different heights from the disc midplane, as the orbit orientations and the positions of particles along their orbits vary. The relative (or impact) velocity varies as well, and with it the collisional outcome. These variations are captured in a Monte Carlo simulation as follows.


We start with a pre-defined grid of orbital elements (the grid to be used in the kinetic model). Each grid bin is defined by its edges and its centre for the eccentricity ($e$), semi-major axis ($a$) (or pericentre distance, $q$) and inclination ($i$). For a total of $N$ grid bins ($N=N_e N_a N_i$), there are $N(N+1)/2$ unique grid bin combinations. In this work limit the grid of semi-major axes to $N_a = 1$, but the methods are general with respect to use of more bins. For each grid bin combination we run a Monte Carlo simulation as follows.

First, we generate two populations of particles for the two sets of orbital elements defining the grid bins. Typically we use $10^4$ particles for the `first' and $10^5$ particles for the `second' population\footnote{This balances the need to increase the accuracy with the need to reduce the computer memory requirements; the latter depends on the number of particles in the `first' population, as explained later.}. The semi-major axes in a population are sampled uniformly from the $a$-bin, while the inclinations and eccentricities are sampled from a very narrow range around the centres of the $e$- and $i$-bins\footnote{This choice reduces artificial diffusion of fragments through the grid. If inclinations and eccentricities are sampled uniformly from their respective bins, very small changes in these orbital elements (including due to small numerical errors) cause particles near bin edges (or the fragments they produce) to `jump' into a neighbouring bin. With limited grid resolution this manifests as numerical diffusion and artificially speeds up the evolution. This artificial diffusion would also occur along the semi-major axis grid, but in this work we only consider a single $a$-bin, as discussed above. Furthermore, the way inclination and eccentricity bins are sampled does not strongly affect the collisional probabilities and velocities.}. 
The rest of the Keplerian orbital elements and the particle locations along the orbit are uniformly distributed in their respective ranges. These populations are used to calculate the fraction of particles in the region of the space where particle orbits overlap, determined radially by particles' pericentres and apocentres and vertically by the smaller of the two inclinations. Then, two new populations of particles are generated in a similar manner, except that they are constrained to the overlapping region. This ensures that the number of particles in the overlapping region remains consistently large to accurately capture collision probabilities and velocities, while the fraction calculated above is accounted for at a later stage. For some parameters, the overlapping region is very narrow and few particles are produced in the overlapping region, which can severely reduce the accuracy of the method; if fewer than one percent of particles fall into the overlapping region, we repeat the calculation with $10^6$ particles. However, the number of particles produced in the second step (in the overlapping region) remains the same (as set on input).

Second, for each particle in the first population we find its nearest neighbour from the other population. As long as the number of particles is large, the choice of the first population here should be unimportant. These pairs of particles represent collisions between the two populations. The finding of the nearest neighbour for each particle is optimised by dividing the space into a grid, and only searching for the nearest neighbour among the particles in the same or the neighbouring grid cell. The size of the grid cell is set to 1\% of the radius of the outer boundary of the overlapping region. If there are no neighbours found for some of the particles, the grid cell size is increased and the process repeated for these particles.

Third, we calculate the collision probability for each pair of particles. The orbit-overlapping region is divided into $N_r=100$ radial ($r$) and $N_\phi=20$ vertical (i.e., polar angle, $\phi$) bins. The fraction of particles in each bin is calculated for each of the populations, and further multiplied by the initial fraction of particles in the entire overlapping region (calculated in the first step). We denote these fractions as $\sigma_1(r)$ and $\sigma_2(r)$. The probability of a collision in each of the bins is then
\begin{equation}
    \Delta(r, \phi) = \frac{\sigma_1(r, \phi) \sigma_2(r, \phi)}{\textrm{d}V(r, \phi)},
\end{equation}
where $dV(r,\phi)$ is the volume of the bin. This quantity is analogous to the $\Delta$-integral calculated analytically in two dimensions by \citet{Krivov2006}. Then, if there are $N_1(r,\phi)$ particles from the `first' population in the given bin, each pair of nearest neighbours identified above is assigned a collision probability
\begin{equation}
    \Delta(j) = \frac{\Delta(r,\phi)}{N_1(r,\phi)}
,\end{equation}
where $j$ counts over all of the particles from the `first' population in this bin (whereas its nearest neighbour from the `second' population need not be from the same $r-\phi$ bin, and can also be the nearest neighbour to more than one particle from the first population).


Furthermore, note that the collision probability assigned in this manner to all $N_1(r,\phi)$ particle pairs is the same, but their relative velocities are not. In each $r$-$\phi$ bin there is a distribution of velocities at which particles collide. This is analogous to how in two dimensions, for a fixed relative orientation of the orbits (or at a fixed radius), particles may collide at up to two different relative velocities \citep{Krivov2006}. The particle pairs are a sample of that velocity distribution.

Finally, for each orbit combination, for each particle in the `first' population we store the components of its position and velocity, the position and velocity of its nearest neighbour from the `second' population, and the probability of their encounter $\Delta(j)$. These data are pre-calculated and stored for use in the kinetic model.

In this work we only consider large particles for which radiation pressure is negligible. If smaller dust particles were considered in the model, the actual velocity of particles in each collision would also depend on their mass, due to radiation pressure. However, in such cases, the Monte Carlo code still needs only to be run once, assuming no radiation pressure for all particles, and the velocities can be scaled using particle $\beta$ parameters when required in the kinetic model (the velocities are simply multiplied by $\sqrt{1-\beta}$). This effect is negligible for large particles and in this work we set $\beta=0$ for all particles, for simplicity. Additionally, in this work the grid of eccentricities for which the Monte Carlo code is run only extends up to the chosen maximum initial eccentricity. That does not need to be the case. If smaller particles were included, which also end up on high-eccentricity orbits because of radiation pressure, the collision rates and velocities would simply be pre-calculated for a grid of eccentricities up to unity. In principle, one could in a similar manner also account for collisions with particles that end up on unbound orbits, but that is not presently implemented.

Furthermore, if very large bodies are considered, the impact velocity calculated from particle velocities from the Monte Carlo simulation should also be corrected for gravitational focusing. The modification to the cross-section is given by $1 + v_{\rm esc}^2/v_{\rm imp}^2$, where $ v_{\rm esc}^2 = 2G(m_{\rm p}+m_{\rm t})/(s_{\rm p}+s_{\rm t}).$ For $M_*=M_\odot$, $r=50$\,au, $s_{\rm t}=s_{\rm p}=1$\,km, $v_{\rm imp}^2 \sim 2 \langle i^2\rangle v_{\rm K}^2$, $\sqrt{\langle i^2\rangle}=10^{-3}$, this correction is only a few percent. Assuming the correction to the particle velocities is similarly small for these parameters, in this work we can safely neglect the role of gravitational focusing.

\subsubsection{Collisional outcomes} \label{sec:methods_outcomes}
Collisions among particles result in catastrophic fragmentation, erosion and/or growth. To determine the outcome of a particular collision, we follow an algorithm employed in other state-of-the-art kinetic models of debris discs \citep[see][]{Lohne2008,Lohne2017} with some minor variations.

The collisional outcome depends on the parameters of the collision (the impact velocity $v_{\rm imp}$ and the impact energy $E_{\rm imp}=m_{\rm t} m_{\rm p} / (m_{\rm t} + m_{\rm p}) v_{\rm imp}^2/2$) and on the properties of the material out of which particles are assumed to be made. The key property of the material here is the critical specific energy for dispersal, $Q^*$, for which we consider 3 different models shown in Table \ref{tab:qdstar} \citep[with the most realistic model based on the results of][]{Benz1999}. 
Following \citet{Stewart2009}, we further define a joint critical specific energy for dispersal of both colliders, $Q^*_{\rm tp}=Q^*(s_{\rm tp})$, where $s_{\rm tp} = (s_{\rm t}^3+s_{\rm p}^3)^{1/3}$.

If the specific impact energy is greater than the joint critical specific energy ($E_{\rm imp} > (m_{\rm t}+m_{\rm p}) Q^*_{\rm tp}$), both colliders are catastrophically disrupted, and their total mass is distributed into a distribution of smaller fragments. The mass of the largest fragment is given by \citep{Paolicchi1996}
\begin{equation}
    m_{\rm lf} = \frac{1}{2} (m_{\rm t}+m_{\rm p}) \left( \frac{E_{\rm imp}}{(m_{\rm t}+m_{\rm p}) Q^*_{\rm tp}} \right)^{-1.24}.
\end{equation}

If the above condition is not fulfilled, colliders lose some mass to erosion. We consider first the case where the target remains largely intact, yet the projectile may still be completely disrupted. We consider this to happen if half of the impact energy is sufficient to disrupt the projectile ($E_{\rm imp}/2 > m_{\rm p} Q^*(s_{\rm p})$). In this case, the target and the projectile are assumed to be eroded as a single body, with the eroded mass given by
\begin{equation} \label{eq:m_frag}
    m_{\rm frag} = \frac12 (m_{\rm t} + m_{\rm p}) \left( \frac{E_{\rm imp}}{(m_{\rm t} + m_{\rm p}) Q^*_{\rm tp}} \right), 
\end{equation}
and the mass of the largest fragment given by $m_{\rm lf} = 0.2 m_{\rm frag}$. The single remnant of such a collision has mass $m_{\rm rem} = m_{\rm t} + m_{\rm p} - m_{\rm frag}$. In some cases, this may lead to net growth of the target \citep[$m_{\rm rem} > m_{\rm t}$;][]{Stewart2009}.

The same outcome happens if the impact velocity is less than the critical sticking velocity (for which we adopt $v_{\rm stick}=1$\,m\,s$^{-1}$). However, note that in such a case the impact energy is so small that very little erosion is expected and the resulting remnant should be larger than either of the colliders.

Finally, if neither of the above conditions is fulfilled, both the target and the projectile are eroded and post-collision their remnants separate. In this case, it is assumed that half of the impact energy is spent on eroding the target and the other half on eroding the projectile. The eroded mass for each is calculated analogous to eq. (\ref{eq:m_frag}), but, using the values of the individual mass of each collider and their individual critical energy for dispersal, and only half of the impact energy. The total eroded mass is distributed into smaller fragments as above.

In this way, every collision results in loss of mass and the mass of the colliders is distributed over the phase-space grid as remnants and as fragments. First, the mass is distributed across the mass bins \citep[we adopt the same power-law mass distribution of fragments as][$n(m)\propto m^{-11/6}$, corresponding to the power-law exponent of $-3.5$ for the distribution of sizes, extending up to $m_{\rm lf}$ and with the normalisation constant such that the total mass of fragments equals the eroded mass]{vanLieshout2014}. Then, orbital elements of these remnants and fragments are determined and distributed over the grid. If a remnant is found to fall into the same bin as its parent body, a fraction of the cratered/accreted mass is moved to a lower/higher mass bin in order to more accurately model the transport of mass in a coarse grid.

A fragment orbit is determined by the location of the collision and the velocities and masses of the colliders (the particle pairs from the Monte Carlo simulation), and also by the effect of radiation pressure on the fragment. The location of the collision is estimated as being the centre-of-mass of the colliders, based on their positions from the Monte Carlo simulation.\footnote{This choice minimises deviation from the conservation of momentum, which arises due to the finite difference in positions of the nearest neighbours in the Monte Carlo simulation.} The location of the collision represents the initial position of the fragments, $\bf{R}$. The initial velocity, $\bf{V}$, of each of the fragments is calculated from conservation of momentum, assuming maximal collisional damping \citep[following][]{Krivov2006,vanLieshout2014},
\begin{equation} \label{eq:cons_mom}
    (m_1+m_2) {\bf V} = m_1 {\bf V}_1 + m_2 {\bf V}_2,
\end{equation}
where $m_1$ and $m_2$ are the masses of the colliders. Velocities of the colliders, ${\bf V}_1$ and ${\bf V}_2$, are the velocities of Monte Carlo particles scaled by their respective $\beta$ parameters to account for the effect of radiation pressure, if radiation pressure is considered.

From the fragment's initial position and velocity the fragment's specific angular momentum $\bf{h}$ is calculated, and finally the orbital elements \citep{Murray2000},
\begin{align}
    a &= \left( \frac{2}{R} - \frac{V^2}{G M_* (1-\beta)} \right)^{-1}, \\
    e &= \sqrt{1 - \frac{h^2}{G M_* (1-\beta) a}}, \label{eq:frag_e}\\
    i &= \textrm{arccos} \left( \frac{h_z}{h} \right).
\end{align}
Here the scaling factor $(1-\beta)$ is added to account for the radiation pressure acting on the fragment; in this work we only consider large bodies for which radiation pressure is negligible and we simply set $\beta=0$ for all particles. 

The orbits of the remnants post-collision are calculated in the same way as for the orbits of the fragments. In other words, every collision results in maximal collisional damping of particle velocities.




\subsection{The phase-space grid} \label{sec:grid}
We use the following grid for the discretisation of the phase space. For the mass we use 26 bins between the (mass corresponding to) particle radius of $s=1$\,mm and $s=100$\,m. The bins are log-spaced such that $m_{i+1}/m_{i} \approx 4$ ($s_{i+1}/s_{i} \approx 1.6$). The bulk density of particles is taken to be 3\,g\,cm$^{-3}$. 

For the eccentricity and the inclination we use 10 linearly spaced bins for each, between zero and $e_{\rm max}$ or $i_{\rm max} = e_{\rm max}/2$; these maximum values differ between models. Lastly, for the semi-major axis we use a single bin between 36 and 44\,au. The numerical model described in this section can be run for multiple semi-major axis bins, in principle. In practice, collisional damping leads to artificially enhanced diffusion of particles through the semi-major axis grid, for the simple fragment distribution method implemented in this model. Since the evolution of semi-major axes is expected to be slow compared to evolution of particle eccentricities and inclinations (see Sect. \ref{sec:theory}), we limit our particles to a single semi-major axis bin.

\section{Results} \label{sec:results}
In all our models, we follow the evolution of a disc around a $M_*=1$\,M$_\odot$ star. All simulations start with a power-law mass distribution ($n(m) \textrm{d}m\propto m^{-11/6} \textrm{d}m$) of bodies $1$\,mm to 100\,m in size (spanning our entire mass grid), placed into a single semi-major axis bin extending from $a_{\rm min}=36$\,au to $a_{\rm min}=44$\,au, with a total mass of the disc $M_{\rm d}=1$\,$M_\oplus$. As discussed in previous sections, the evolution of the semi-major axes of 
particles due to collisional damping is neglected in this work. These model parameters have been chosen so that even the largest bodies damping timescales (calculated using eq. (\ref{eq:r_damp})) are shorter than 1\,Gyr, which is for how long we run all our simulations. 

\begin{figure}
    \centering
    \includegraphics[width=\columnwidth,clip,trim={0 0 0 1.3cm}]{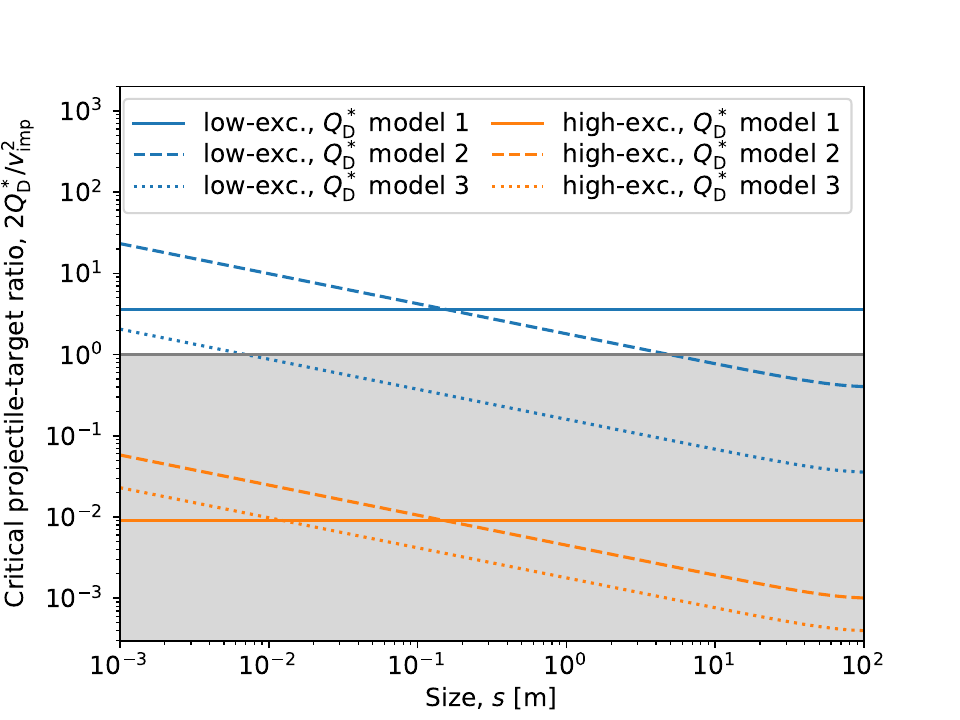}
    \caption{Approximate initial critical projectile-to-target mass ratio ($Y_{\rm c}$) for the models considered in Sects. \ref{sec:qdstar_const} and \ref{sec:qdstar_size_vel}. Blue and orange lines are for low-excitation ($e_{\rm max}=0.01$) and high-excitation discs ($e_{\rm max}=0.2$), respectively, and the different line styles are for different $Q_{\rm D}^*$ models shown in Table \ref{tab:qdstar}; see plot legend. Here for the average impact velocity we adopt $v_{\rm imp}=v_{\rm K} e_{\rm max}/2$, evaluated at the centre of the belt, at 40\,au. The nearly power-law size-dependence is due to only considering sizes up to 100\,m, which are all in the strength regime of material strength. Where the mass ratio shown is higher than unity, the exact value of it is unphysical, as the assumptions behind the simple formula break down, as discussed in Sect. \ref{sec:theory}. Nevertheless, this figure shows for which parameters we expect the disc evolution to be dominated by fragmentation, as opposed to bouncing or cratering.}
    \label{fig:crit_ratio}
\end{figure}

The eccentricities and inclinations of the particles are initially uniformly distributed between zero and $e_{\rm max}$ and $i_{\rm max}=e_{\rm max}/2$, respectively. We consider two sets of values for these parameters: a high-excitation disc with $e_{\rm max}=0.2$ and a low-excitation disc with $e_{\rm max}=0.01$. Together with the rest of the model parameters, these two sets of values ensure that we probe both systems in which the damping rate is faster and slower than the fragmentation rate, or in other words, both systems in which the critical projectile-to-target mass ratio is larger and smaller than unity. For both low-excitation and high-excitation discs we further consider three different, increasingly realistic models for the specific critical disruption energy, $Q_{\rm D}^*$, shown in Table \ref{tab:qdstar}. First, for simplicity, we consider a constant value (model 1), then we consider a more realistic size-dependent function (model 2), and, lastly, the full size and velocity-dependent function (model 3). In Fig. \ref{fig:crit_ratio} we illustrate how the approximate initial critical projectile-to-target mass ratio ($Y_{\rm c}=2 Q_{\rm D}^* / v_{\rm imp}^2$) varies with particle size in these 6 different disc models. This approximate critical ratio aids the discussion of our results throughout this section, although it is not directly used in our kinetic model (for the description of how collisional outcomes are implemented in the model, see Section \ref{sec:methods_outcomes}).

We present our results in the following order. First, in Sect. \ref{sec:inbounc}, as a test of our model, we consider a disc in which particles can only bounce off each other, inelastically, but without any cratering or fragmentation. Then, in Sects. \ref{sec:qdstar_const} and \ref{sec:qdstar_size_vel} we consider a collisional cascade with fragmentation and cratering, with different excitation levels and with different, increasingly realistic, material strength prescriptions, as outlined above.

\subsection{Inelastic bouncing of indestructible particles} \label{sec:inbounc}

\begin{figure*}[!]
    \centering
    \includegraphics[width=\columnwidth]{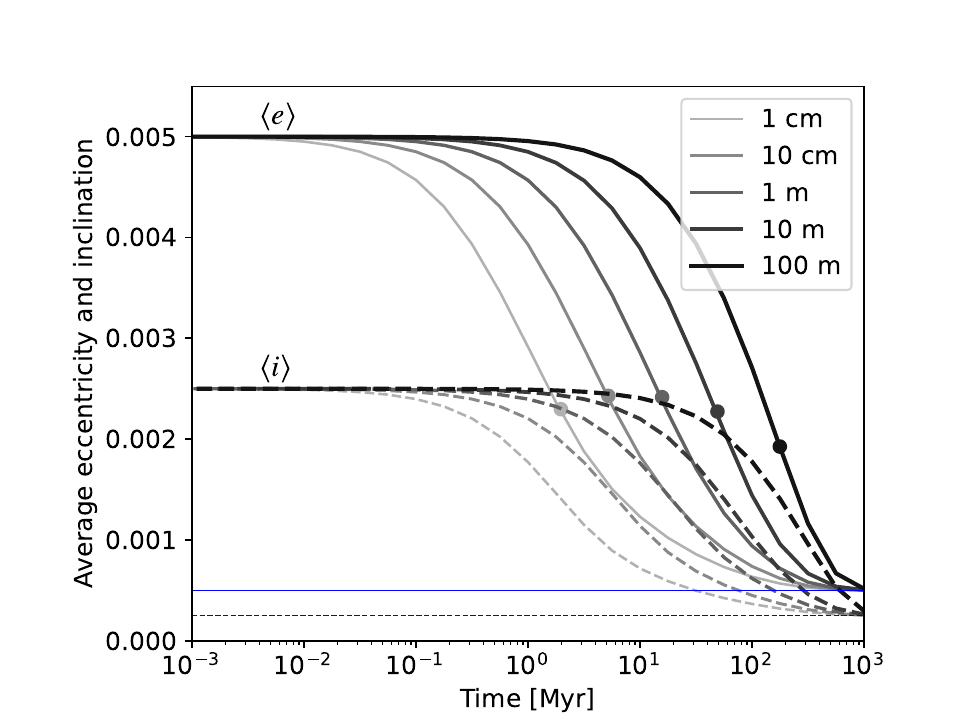}
    \includegraphics[width=\columnwidth]{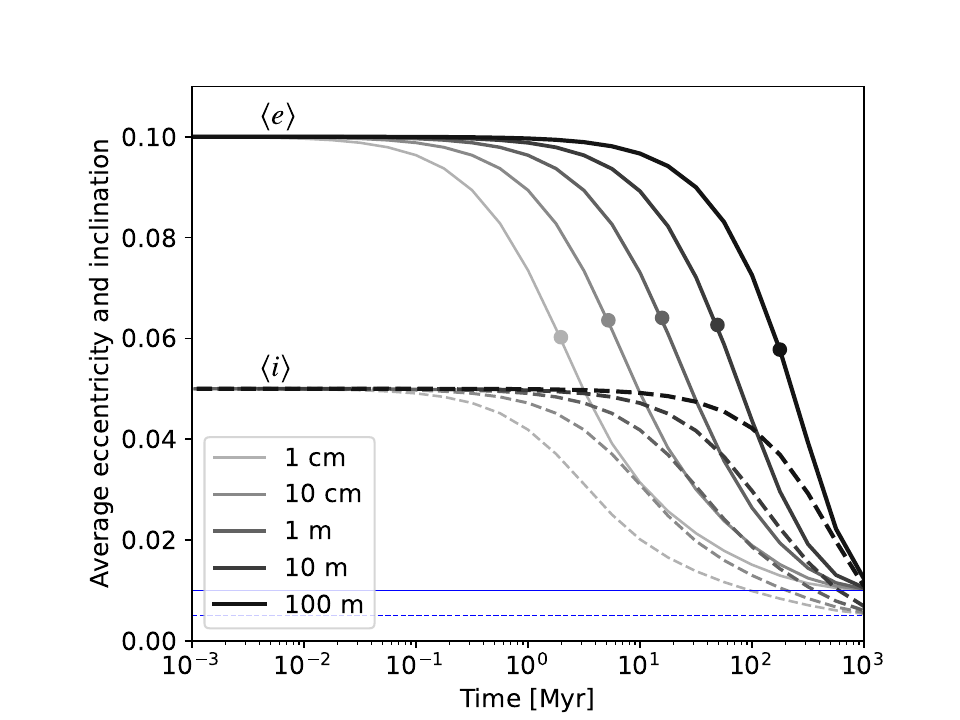}
    \caption{Evolution of average eccentricity (solid lines) and inclination (in radians; dashed lines) in a disc of indestructible particles, inelastically bouncing off each other in collisions. The left- and right-hand-side panels show a low-excitation disc and high-excitation disc, respectively. Horizontal blue lines show the lowest grid bin values, i.e. the lowest values to which eccentricity and inclination can be damped in the model. Circles on solid curves indicate damping timescales predicted by a simple particle-in-a-box model. See Sect. \ref{sec:inbounc}.}
    \label{fig:inbounc}
\end{figure*}

To test our numerical model, we first consider a disc of indestructible particles. Each collision leads to complete damping of the relative velocity between particles, but the particles remain whole and separate, there is no mass evolution. As expected, this leads to damping of the average eccentricity and inclination, as shown in Fig. \ref{fig:inbounc}. Particles are only damped down to the lowest-value grid bins of the kinetic model, and so the damping is limited artificially \citep[this is an unfortunate consequence of discretisation which is present in other work as well, e.g.][]{Brahic1977,Lithwick2007}. As the average eccentricities and inclinations approach these artificial lowest values, the damping slows down. To some degree this is also due to discretisation (see Fig. \ref{fig:test_res}), but for the smaller bodies there is also a physical reason for this - collisional stirring by the larger bodies whose eccentricities and inclinations are damped more slowly. Overall, the model results in damping by a factor of $10$ within 1\,Gyr.

\begin{table}
\caption{Models we consider for the specific critical disruption energy, $Q_{\rm D}^*$.}
\label{tab:qdstar}
\begin{center}
\begin{tabular}{ c l }
  \hline
  model & $Q_{\rm D}^*$ \\ \hline
  1 & $10^7 \frac{\textrm{erg}}{\textrm{g}}$ \\ 
  2 & $5 \times 10^6 \frac{\textrm{erg}}{\textrm{g}} \left( \left( \frac{s}{\textrm{1\,m}} \right)^{-0.37} + \left( \frac{s}{\textrm{1\,km}} \right)^{1.38} \right) $ \\  
  3 & $5 \times 10^6 \frac{\textrm{erg}}{\textrm{g}} \left( \left( \frac{s}{\textrm{1\,m}} \right)^{-0.37} + \left( \frac{s}{\textrm{1\,km}} \right)^{1.38} \right) \left( \frac{v_{\rm imp}}{\textrm{3\,km/s}} \right)^{0.5}$ \\
  \hline
\end{tabular}
\end{center}
\end{table}

We can compare these results to a damping timescale from a simple particle-in-a-box model, that is, to the inverse of rate $R_{\rm damp}$ given by eq. (\ref{eq:r_damp}). In Fig. \ref{fig:inbounc}, the particle-in-a-box damping timescales for different particle sizes are shown as filled circles. Particle eccentricities fall off by a factor of $\sim 2$ in a single particle-in-a-box damping timescale, and it takes about 10 times longer than the damping timescale to completely damp eccentricity and inclination for any given particle size; this is in line with results from N-body simulations \citep[][]{Brahic1977,Lithwick2007}. We do not make a direct, quantitative comparison with these studies because of differing assumptions (complete damping is assumed in our model, while the N-body simulations assume nonzero elasticity).

Far above our artificial size distribution cut-off at the size of 1\,mm, the particle-in-a-box damping timescale depends on the particle size as $\propto s^{1/2}$ (see Sect. \ref{sec:theory}). The results of our kinetic model agree well with this in both the low-excitation and the high-excitation disc. However, there is a factor of $\sim 2$ difference in the damping timescale in the high-excitation and the low-excitation numerical model (the left and the right-hand-side panel of Fig. \ref{fig:inbounc}). This is contrary to the simple particle-in-a-box estimate of the damping rate, which does not depend on the level of excitation in the disc. We ascribe this to over-simplification in the particle-in-a-box model. In obtaining eq. (\ref{eq:r_coll_constant}), both the volume of interaction and the relative velocity are assumed to be proportional to the average inclination, which then cancels out. In the kinetic model (and in reality), both quantities depend in a more complex way on both eccentricity and inclination. For example, in the particle-in-a-box model we set the disc width to be the width of our semi-major axis bin, whereas particle pericentres and apocentres are outside of that range, and depend on particle eccentricities. 


\subsection{Collisional cascade with fixed material strength} \label{sec:qdstar_const}
Here and in the following section, we consider collisional cascades where the mass of particles in each size bin evolves due to fragmentation, cratering, and, if velocities are sufficiently damped, growth. In this section, we assume a constant specific critical disruption energy $Q_{\rm D}^*$ for all particles (model 1 in Table \ref{tab:qdstar}). We consider the full realistic $Q_{\rm D}^*$ prescription (models 2 and 3 in Table \ref{tab:qdstar}) in the following section.

\begin{figure*}[h]
    \centering
    \includegraphics[width=\columnwidth]{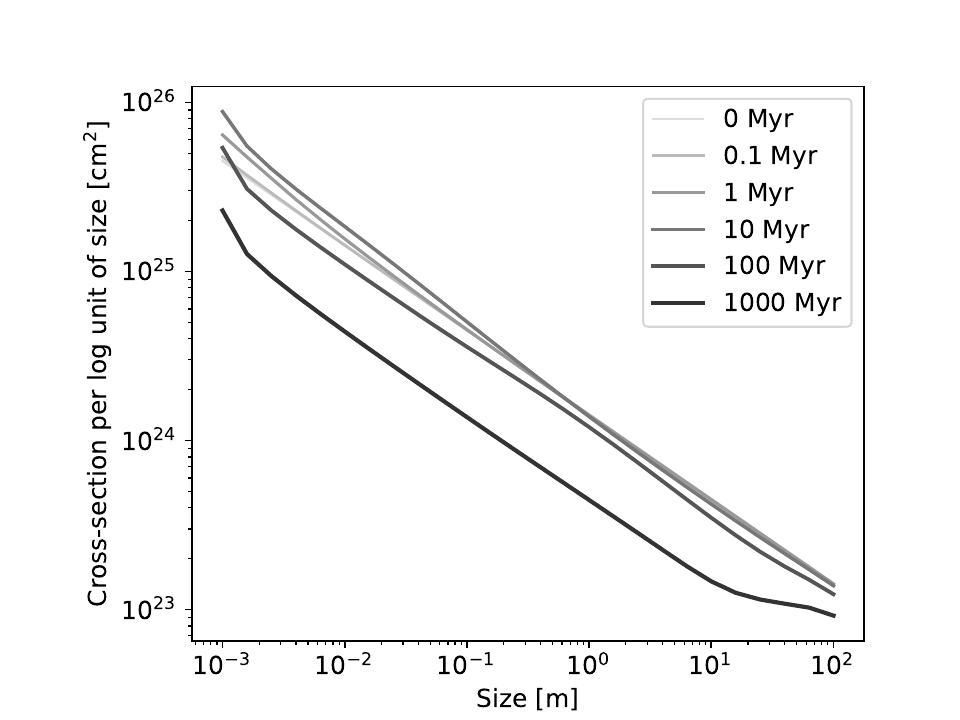}
    \includegraphics[width=\columnwidth]{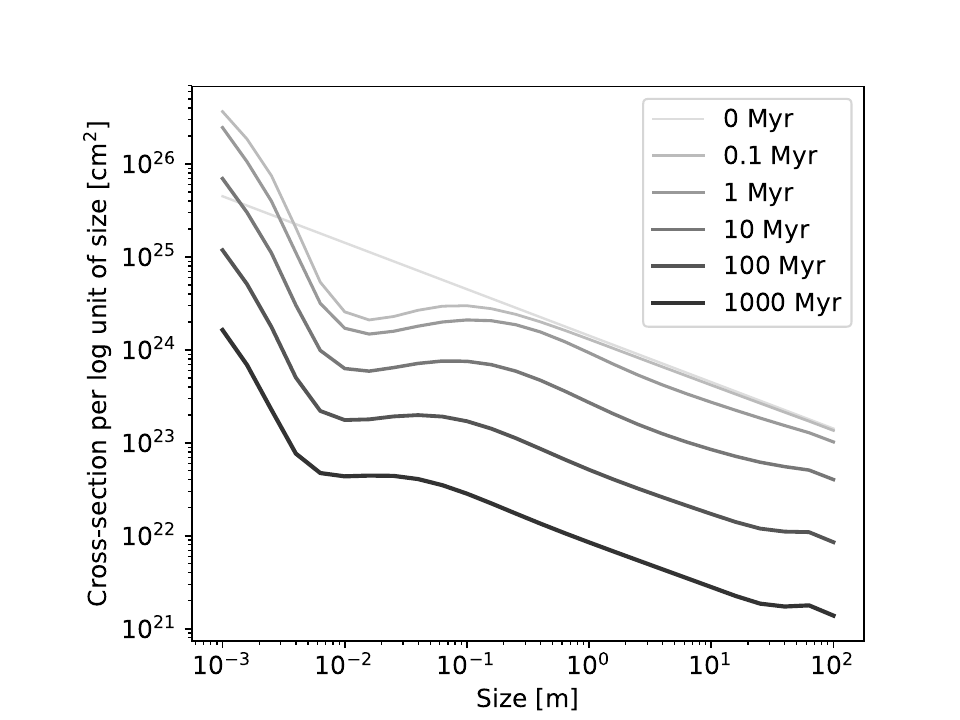}
    \caption{Cross-section area per base-10 logarithmic unit of size at different times (see plot legend) in a model of a collisional cascade with constant $Q_{\rm D}^*$ (model 1 in Table \ref{tab:qdstar}). The left- and right-hand-side panels show a low-excitation disc and high-excitation disc, respectively. See Sect. \ref{sec:qdstar_const}.}
    \label{fig:qdstar_const_crosec}
\end{figure*}

The overall mass evolution of our collisional cascades agrees well with previous analytic and numerical work \citep[e.g.][see Appendix \ref{sec:tests_km} for a more detailed comparison]{Dominik2003,Krivov2006,Thebault2007,Wyatt2007a}. The time-evolving size distributions for the low-excitation and the high-excitation models are shown in Fig. \ref{fig:qdstar_const_crosec}. On the collisional timescale of largest bodies a steady-state is established, and after that the size distribution shape varies very little. The total mass in the system decreases steadily with time, as the mass in the largest bodies is ground down to the lowest sizes, below which it is considered to be lost. In reality mass is lost when particles are fragmented below the radiation pressure blowout size. In our models the lower-size cutoff is set instead to 1\,mm. This should not affect the evolution of the disc model overall, although it does introduce an artificial over-abundance of 1\,mm grains, as there are no smaller projectiles to collide with for them. This is analogous to the over-abundance of grains just above the blowout size limit that is present when modelling the full size distribution \citep[e.g.][]{CampoBagatin1994,Krivov2006,vanLieshout2014}. 


\begin{figure*}
    \centering
    \includegraphics[width=\columnwidth]{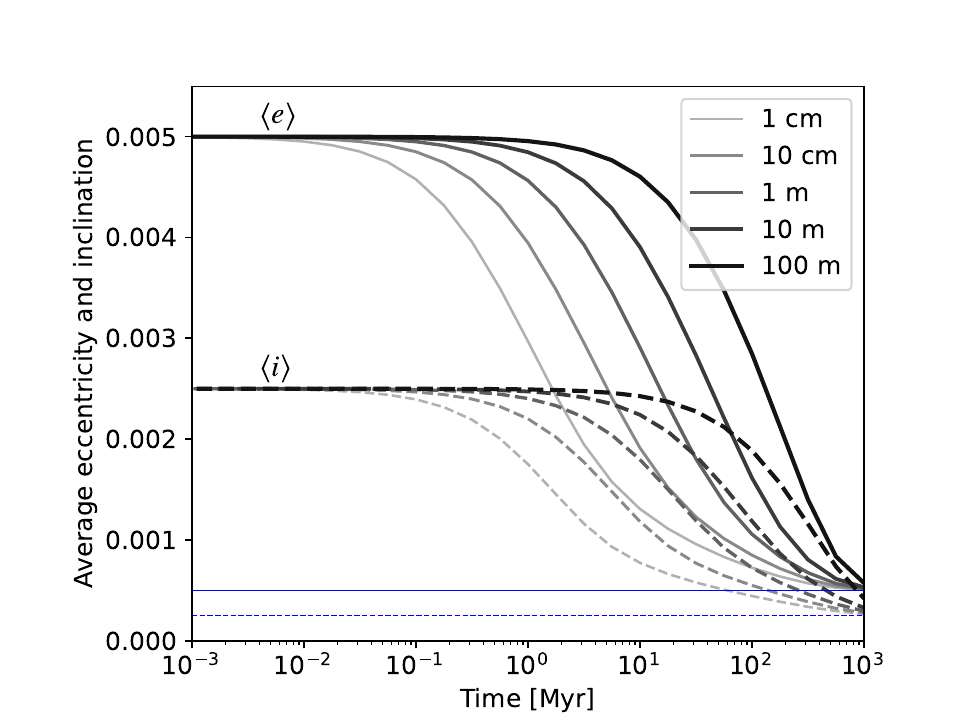}
    \includegraphics[width=\columnwidth]{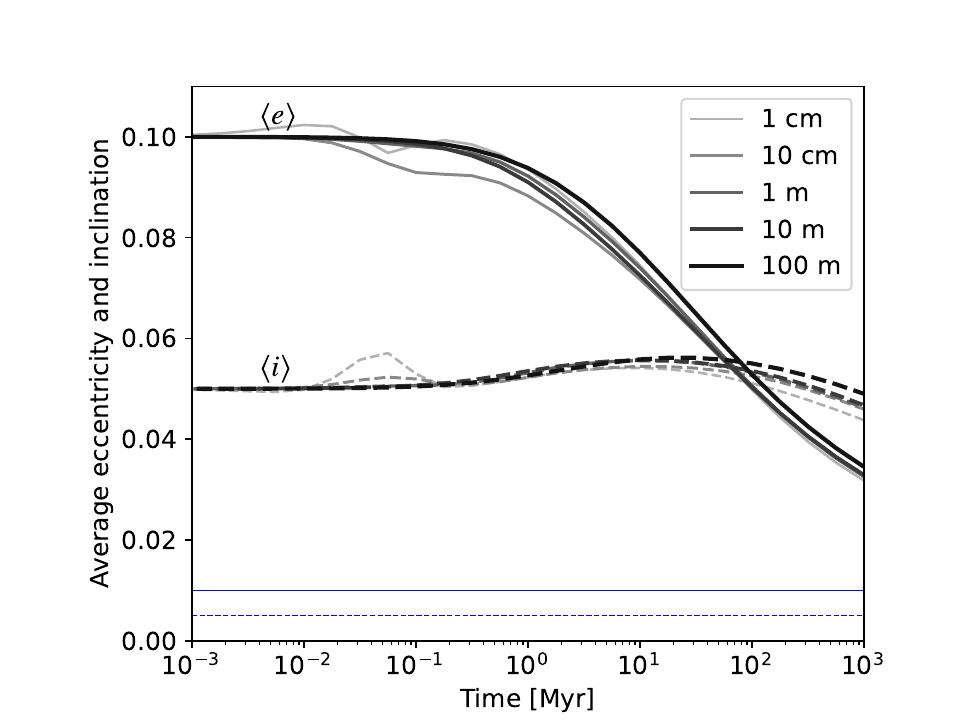}
    \caption{Evolution of average eccentricity (solid lines) and inclination (in radians; dashed lines) in a model of a collisional cascade with constant $Q_{\rm D}^*$ (model 1 in Table \ref{tab:qdstar}). The left- and right-hand-side panels show the low-excitation disc and high-excitation disc, respectively. Horizontal blue lines show the lowest grid bin values, i.e. the lowest values to which eccentricity and inclination can be damped in the model. See Sect. \ref{sec:qdstar_const}.}
    \label{fig:qdstar_const}
\end{figure*}

\begin{figure*}
    \centering
    \includegraphics[width=\columnwidth]{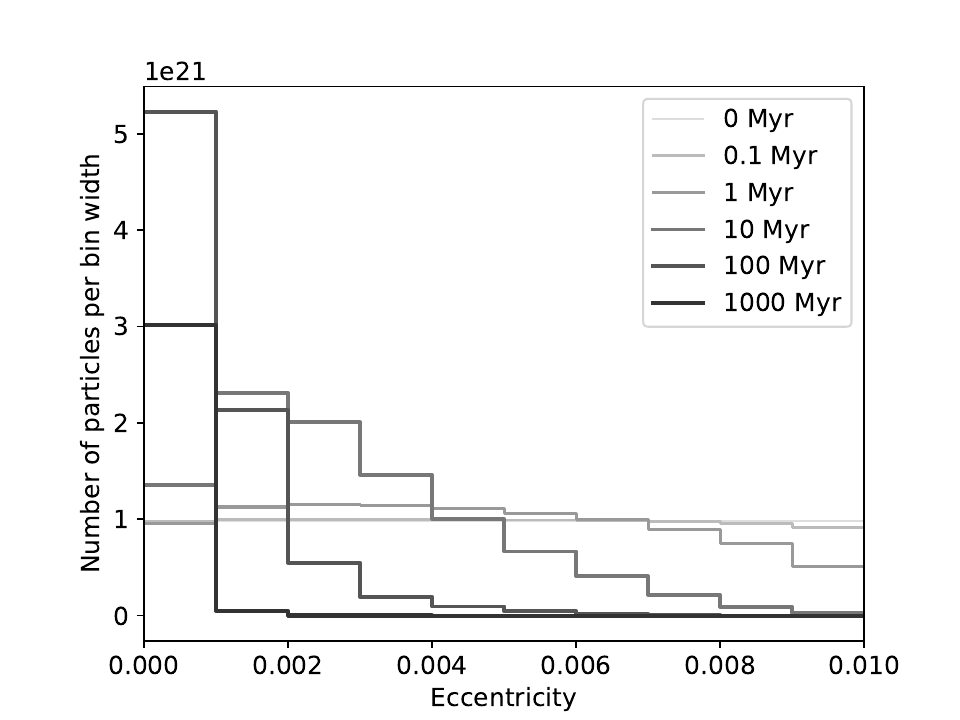}
    \includegraphics[width=\columnwidth]{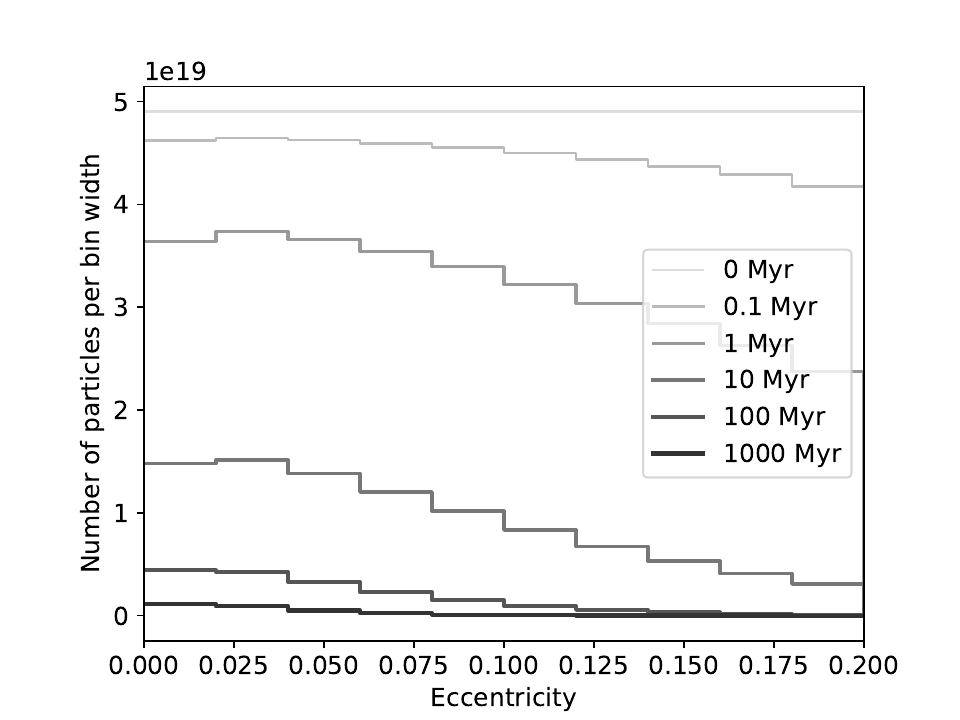}
    \includegraphics[width=\columnwidth]{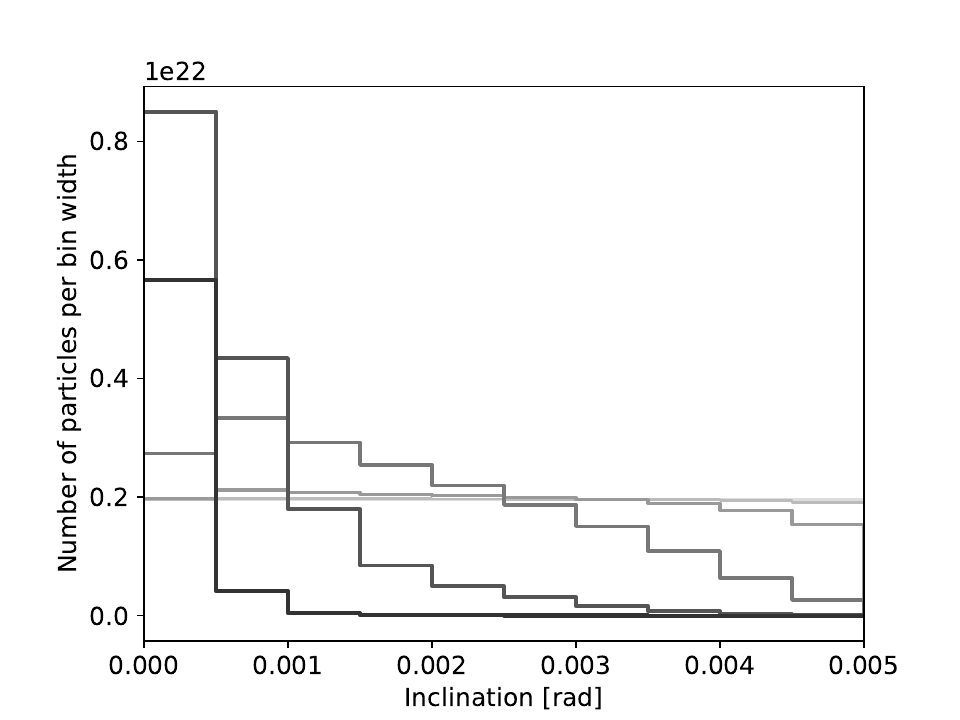}
    \includegraphics[width=\columnwidth]{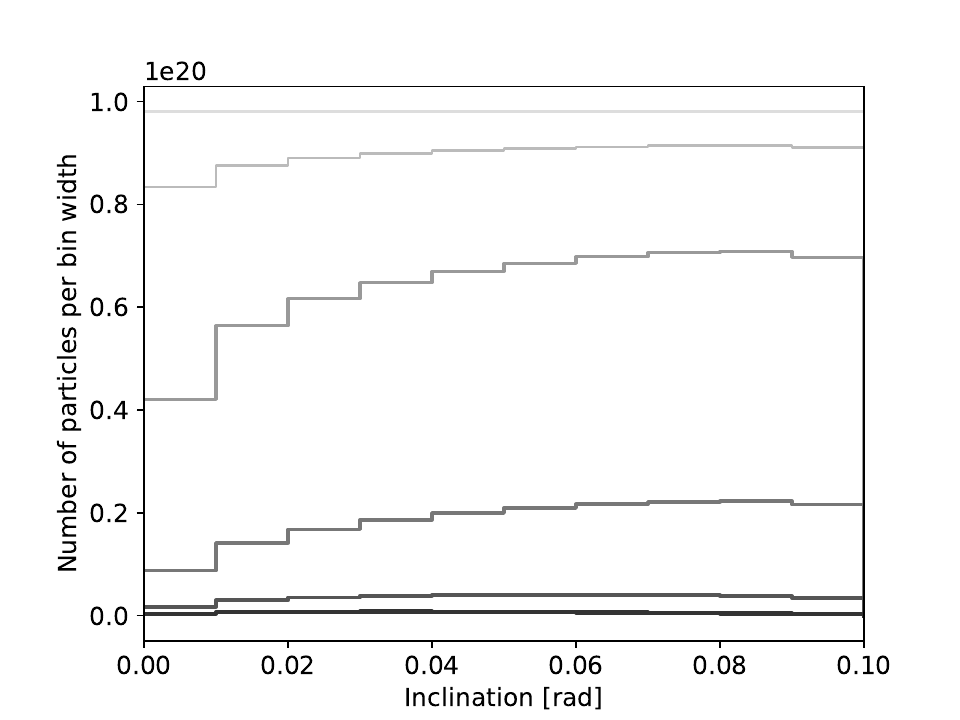}
    \caption{Evolution of eccentricity (top panels) and inclination (bottom panels) distributions for 1\,m bodies in the models shown in Figs. \ref{fig:qdstar_const_crosec} and \ref{fig:qdstar_const}. The left- and right-hand-side panels show low-excitation disc and high-excitation disc, respectively. See Sect. \ref{sec:qdstar_const}.}
    \label{fig:qdstar_const_distr}
\end{figure*}

The evolution of the average eccentricities and inclinations for particles of various sizes are shown in Fig. \ref{fig:qdstar_const} for the low-excitation and the high-excitation disc model, in the left-hand-side and right-hand-side panels, respectively. The results are entirely consistent with the basic theory laid out in Sect. \ref{sec:theory}. For the low-excitation disc, the approximate critical projectile-to-target mass ratio evaluates to $Y_{\rm c} \sim 4$ for the approximate average impact velocity (see Fig. \ref{fig:crit_ratio}). Indeed, in the low-excitation disc, eccentricities and inclinations are efficiently damped, on practically the same timescales as in our inelastic-bouncing models (compare with left-hand-side panel of Fig. \ref{fig:inbounc}). Looking at the distributions of eccentricity and inclination (see left-hand-side column in Fig. \ref{fig:qdstar_const_distr}), we also find that mass is redistributed into lower-$e$ and lower-$i$ bins faster than it is lost to destructive-collisional evolution of the disc.

Conversely, in the high-excitation disc the eccentricities and the inclinations are not damped by the end of the simulation, and the inclinations in particular change very little over the course of 1\,Gyr. The initial approximate critical projectile-to-target mass ratio is $Y_{\rm c} \sim 10^{-2}$ for this disc (see Fig. \ref{fig:crit_ratio}). In this case, distributions of eccentricity and inclination are shaped by destructive collisions. The right-hand-side top and bottom panel of Fig. \ref{fig:qdstar_const_distr} show how the number of 1\,m bodies monotonously decreases with time in all eccentricity and inclination bins. The decrease is faster for higher eccentricities and for lower inclinations, consistent with the damping of the average eccentricity and the slight increase in the average inclination seen in Fig. \ref{fig:qdstar_const}; the reasons for this are discussed in detail in the rest of this section. The results are also remarkably independent of particle size.

\begin{figure}
    \centering
    \includegraphics[width=\columnwidth]{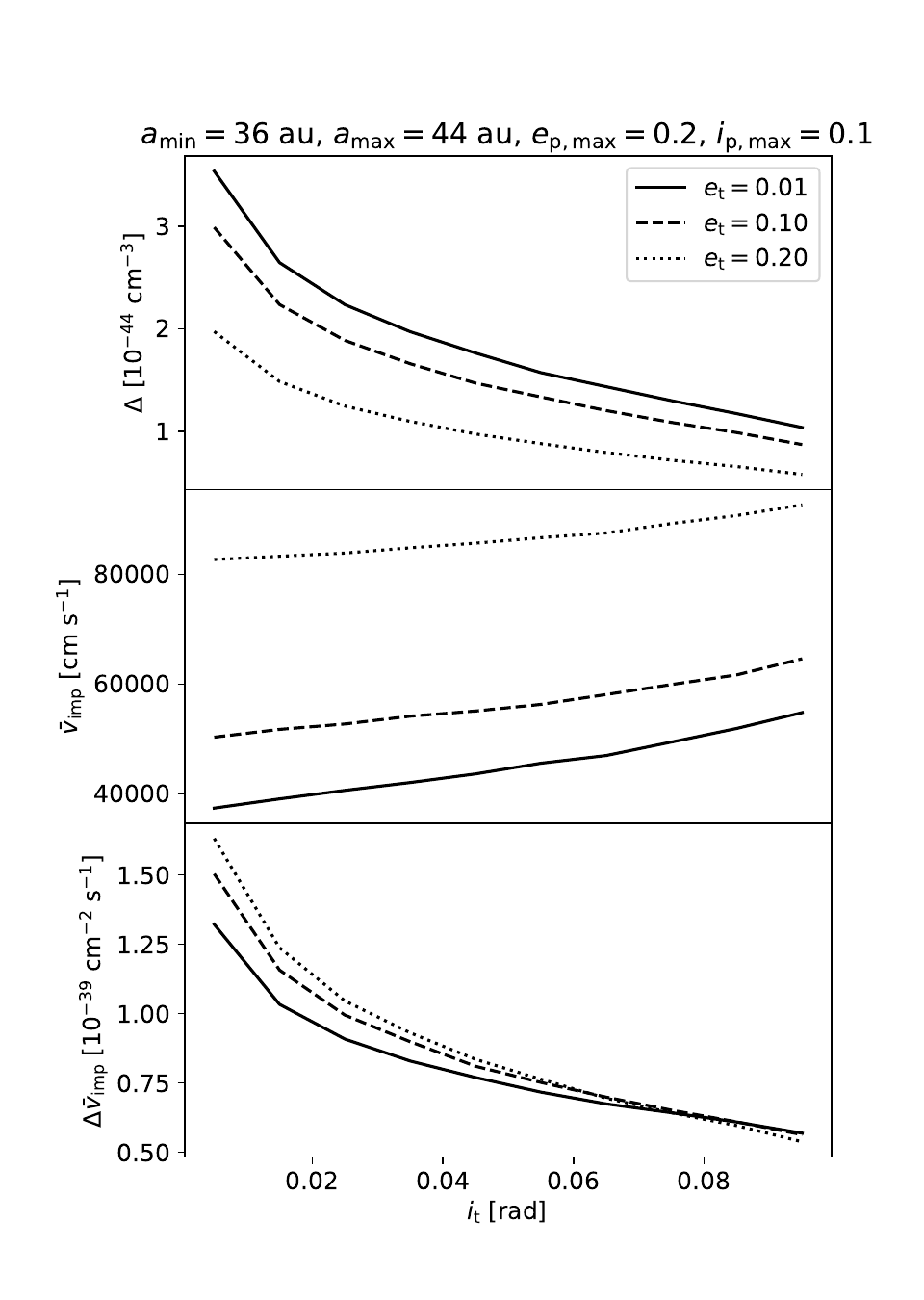}
    \caption{Geometric collision probability (top), average impact velocity (middle), and the geometric part of the collision rate (bottom) in the high-excitation simulation discussed in Sect. \ref{sec:qdstar_const}, as a function of the eccentricity $e_{\rm t}$ and the inclination $i_{\rm t}$ of target particles colliding with projectiles with uniform distributions of eccentricity and inclination from 0 to $e_{\rm p, max}=0.2$ and $i_{\rm p, max}=0.1$, respectively, calculated using the Monte Carlo code described in Sect. \ref{sec:methods_mc}. Both targets and projectiles have uniform distributions of semi-major axis with $a_{\rm min}=36$\,au and $a_{\rm max}=44$\,au, as in our disc models. In these simulations, we use $10^5$ particles in each Monte Carlo population. The collision rate is maximised for high eccentricities and low inclinations of the target. See Sect. \ref{sec:qdstar_const}.}
    \label{fig:qdstar_const_rate}
\end{figure}

The different evolution of eccentricities and inclinations can be explained by how the destruction rate depends on a particle's eccentricity and inclination. We illustrate this using the Monte Carlo code described in Sect. \ref{sec:methods_mc}. We calculate the geometric part of the collision rate of a target on orbits of varying eccentricity and inclination with a population of projectiles with uniform distributions as in the initial conditions of our kinetic models. Figure \ref{fig:qdstar_const_rate} shows how the geometric collision probability (top panel) mainly depends on the target inclination, peaking for targets on low-inclination orbits, moving through the densest region of the disc near midplane, and the impact velocity (middle panel) mainly depends on target eccentricity. The geometric collision probability also decreases with increasing eccentricity of the target, because the particles on high-eccentricity orbits spend considerable time outside of the high-density centre of the belt (for these particular parameters, for which the width of the belt in terms of semi-major axes, $a_{\rm max}-a_{\rm min}$, is comparable to the radial width of high-eccentricity orbits being considered). Nevertheless, the geometric part of the collision rate (bottom panel) peaks for high eccentricities and for low inclinations of the target. The destruction rate, as opposed to the collision rate shown here, has further dependence on the velocity, as only collisions above the critical impact velocity will result in fragmentation. This further amplifies the loss rate for particles on high-eccentricity orbits. Overall, thus, Fig. \ref{fig:qdstar_const_rate} shows why in our high-excitation disc the average eccentricity decreases with time, and why the average inclination initially increases. At later times, the average inclination is also decreasing with time, perhaps because some of these trends in how the destruction rates depend on inclination change after the eccentricities are damped to some degree. In general, as the eccentricity and inclination distributions of both targets and projectiles change over time so do the collision and destruction rates of objects on different orbits. Because of this, our results may be sensitive to our initial conditions and different initial distributions should be explored in future work.

Destructive collisions do not only destroy particles, they also produce them (i.e., they produce fragments and remnants). We find that the production of particles typically has the opposite effect from destruction: if destruction damps eccentricities of bodies of a particular size, production will typically increase them. This is likely a consequence of the fragments inheriting the velocity distribution of their larger parent bodies: the parent bodies are also being preferentially destroyed at higher eccentricities, producing fragments at higher eccentricities. Nevertheless, through inspection of loss and gain rates in our kinetic model we find that the destruction dominates the change rates of the average eccentricity and inclination, although the effects are of the same order of magnitude. This is to be expected in a collisional cascade, where the disc is losing mass and for each size bin the loss rate is at least somewhat higher than the gain rate. An additional possible reason for this imbalance in the evolution of the average eccentricities may be that bodies are indeed inheriting the eccentricity distribution of the destroyed larger bodies, but not exactly. Due to collisional damping the newly created particles should have at least somewhat lower eccentricities than the destroyed ones. However, given the small projectile-target mass ratio in our high-excitation disc, most fragments will simply have inherited a velocity distribution very close to that of the massive target.

Another key feature of how the high-excitation disc evolves is that the ratio of the average eccentricity to the average inclination drops from the initial factor of 2 to a bit below unity. This is contrary to findings of, for example, \citet{Brahic1977}, where the system that initially has a different eccentricity-to-inclination ratio evolves towards the value of $\sim 2$. This is a simple consequence of our assumption of complete collisional damping. With zero elasticity in physical collisions \citep[and also from not having any gravitational interactions as in, for example,][]{Ida1992}, there is nothing pushing the system towards energy equipartition.

\begin{figure*}
    \centering
    \includegraphics[width=\columnwidth]{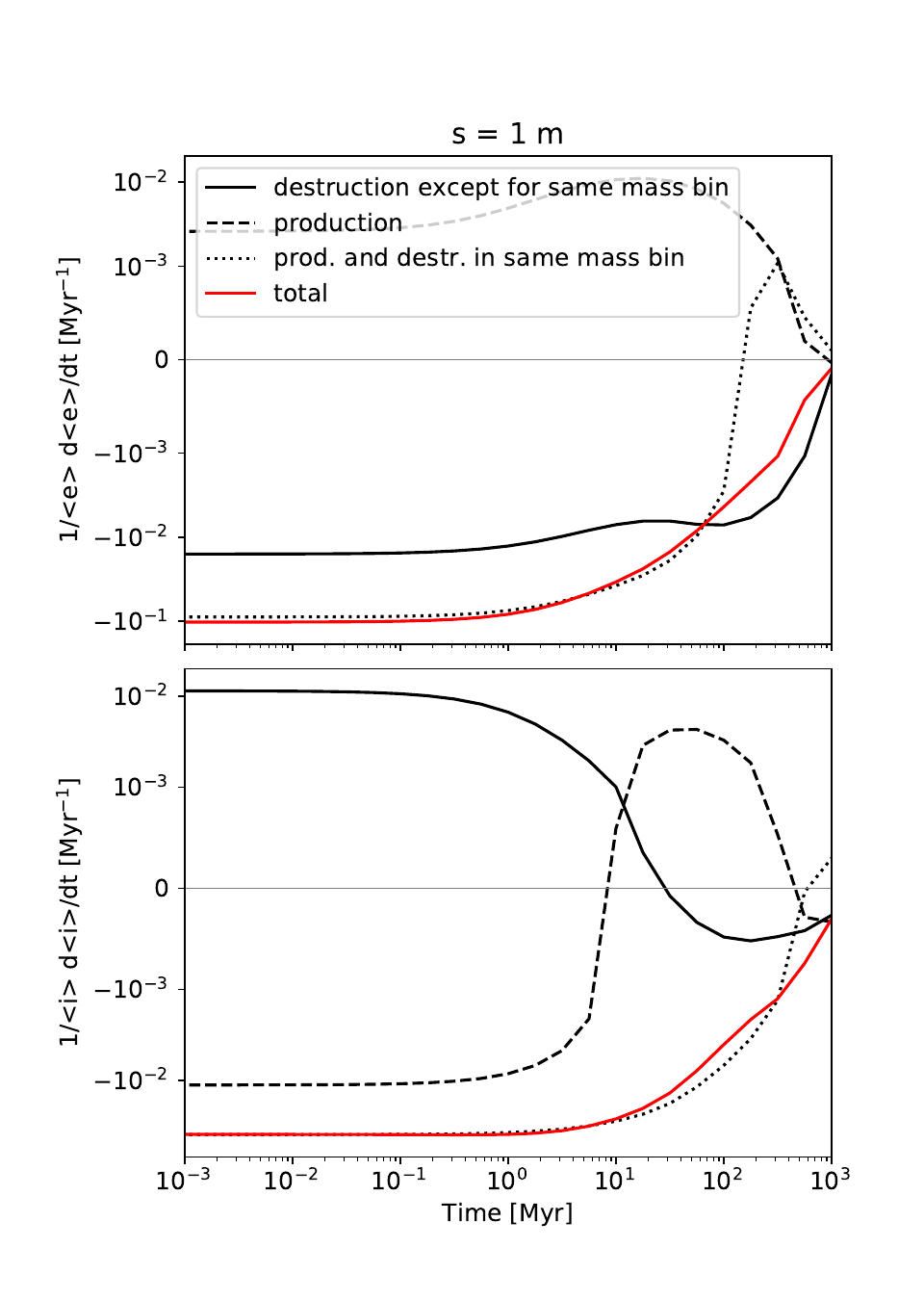}
    \includegraphics[width=\columnwidth]{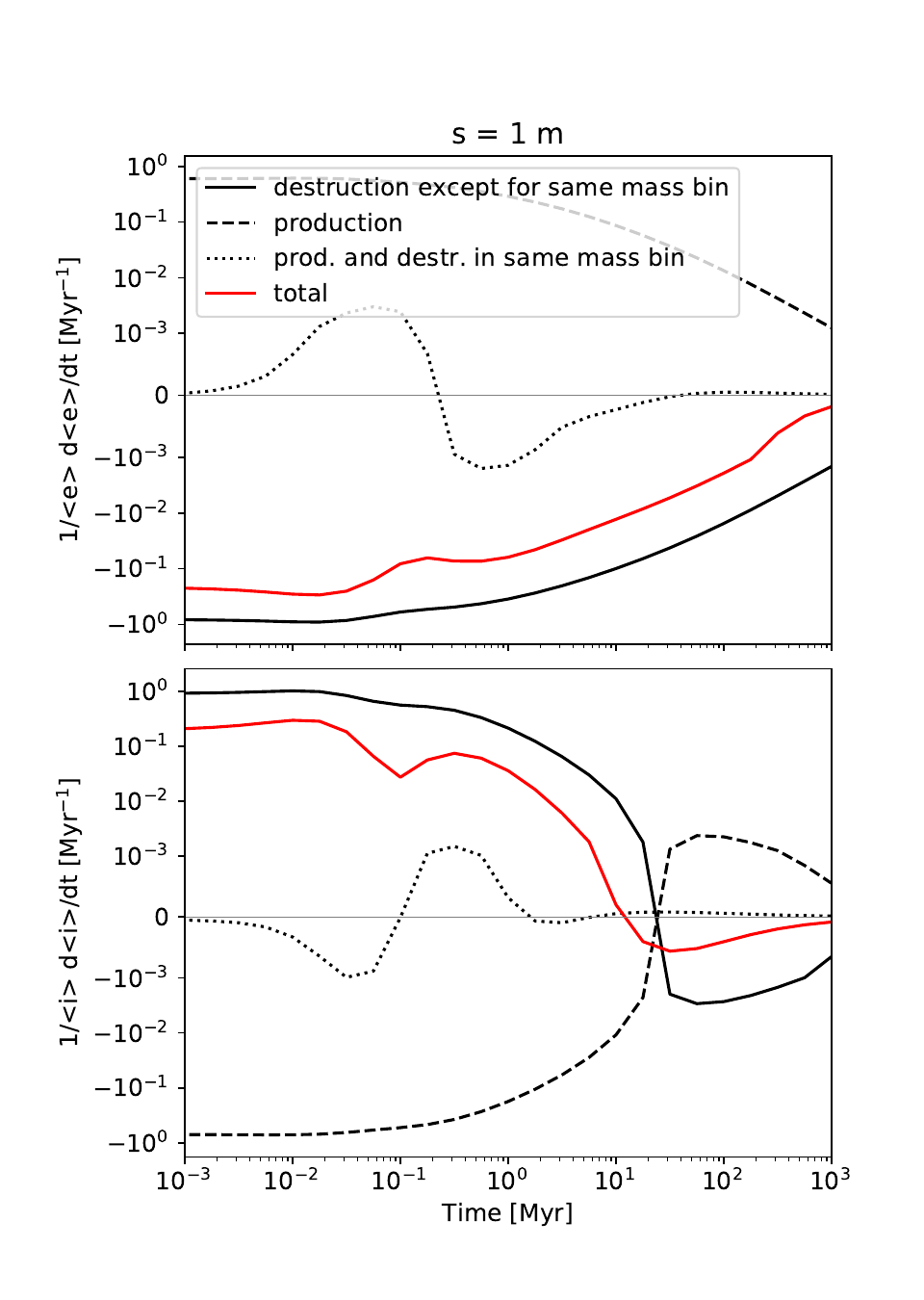}
    \caption{Rate of change of average eccentricity (top panels) and inclination (bottom panels) for 1\,m bodies in the models shown in Figs. \ref{fig:qdstar_const_crosec} and \ref{fig:qdstar_const}. The left- and right-hand-side panels show low-excitation disc and high-excitation disc, respectively. The total rate of change is shown in red colour, while the black lines show how different types of collisions contribute to the total rate (see plot legend and the explanation of the classification of collisions in Sect. \ref{sec:qdstar_const}).}
    \label{fig:qdstar_const_diag}
\end{figure*}

We can further confirm our interpretation of these results by looking into what types of collisional outcomes govern the evolution of the average eccentricity and inclination. In our kinetic model phase space the average eccentricity as a function of particle size is calculated as
\begin{equation}
    \langle e \rangle (j_m) = \frac{\sum_{j_e, j_i} n(j_m, j_e, j_i) e(j_e)}{\sum_{j_e, j_i} n(j_m, j_e, j_i)},
\end{equation}
where the denominator equals $n(j_m)$, the number of particles in mass bin $j_m$, and $j_e$ and $j_i$ are eccentricity and inclination bin indices. Its change rate is given by
\begin{equation}
    \frac{\textrm{d}\langle e \rangle(j_m)}{\textrm{d}t} = \frac{1}{n(j_m)} \sum_{j_e,j_i} \frac{\textrm{d}n(j_m,j_e,j_i)}{\textrm{d}t} \left( e(j_e) - \langle e \rangle(j_m) \right).
\end{equation}
Analogous equations can be written down for the average inclination. Here we can separate the change rate of the number of particles, $\textrm{d}n(j_m,j_e,j_i)/\textrm{d}t$, into contributions from different collisional outcomes. However, it is technically difficult to separate this into the exact collisional outcomes itemised in Sect. \ref{sec:methods_outcomes}. After the pre-calculation phase in our code, during time evolution, this information is not kept, to reduce computational costs. Instead, we can separate the change rate into:
\begin{enumerate}[(1)]
    \item the rate at which particles of size $j_m$ are produced in collisions in which neither of the colliders is from the same size bin (this is production through fragmentation or cratering),
    \item the rate at which mass is removed from size bin $j_m$ and moved to a different size bin (this is destruction of particles), and
    \item the rate at which mass is removed from and re-added to size bin $j_m$ in collisions where at least one of the colliders is from the same size bin (i.e. mass is being moved across the eccentricity and inclination grid, but not the mass grid).
\end{enumerate}
We show the results of this analysis in Fig. \ref{fig:qdstar_const_diag}. As expected, in the high-excitation disc, the change rate is determined by the balance between pure destruction and production of particles by fragmentation and cratering.

In the low-excitation model the change rate of both eccentricity and inclination is governed by the third `category' of collisions. These collisions are associated with some of the mass remaining in the same mass bin. This is in line with our theoretical expectations that collisions in low-excitation discs mostly result in cratering/bouncing, with efficient collisional damping. However, the rates included in category 3 on Fig. 7 also include collisions in which a projectile from the size bin $j_m$ is completely destroyed in a collision with a larger body, but then a fragment of the same size is created in that collision (i.e. a part of the larger body). We cannot differentiate between those two collisional outcomes at runtime in our simulations. 

\subsection{Collisional cascade with size- and velocity-dependent material strength} \label{sec:qdstar_size_vel}
\begin{figure*}
    \centering
    \includegraphics[width=\columnwidth]{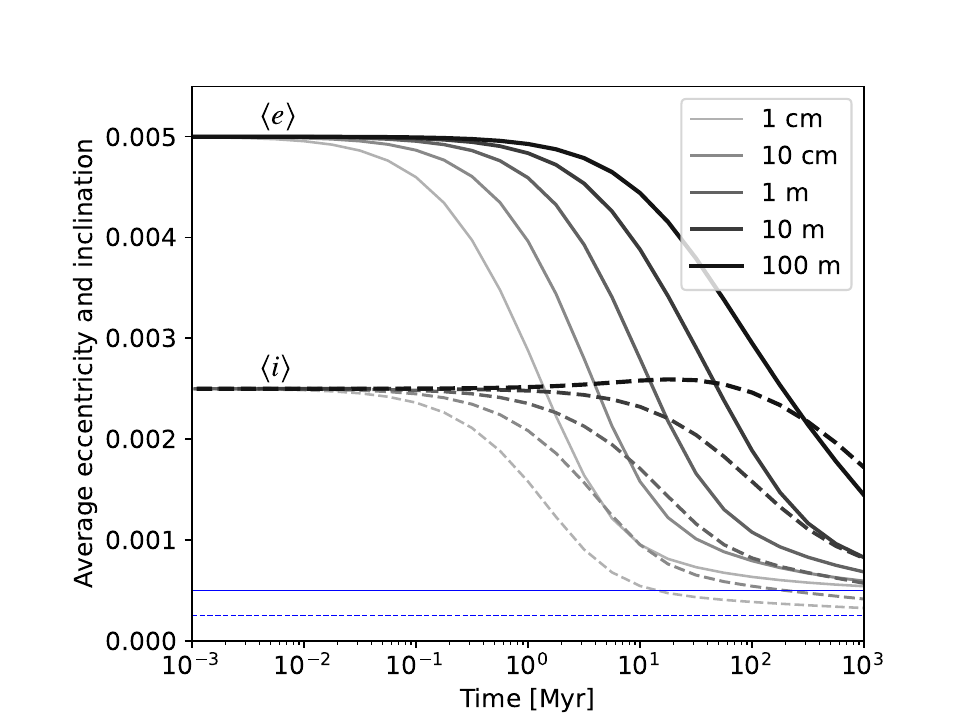}
    \includegraphics[width=\columnwidth]{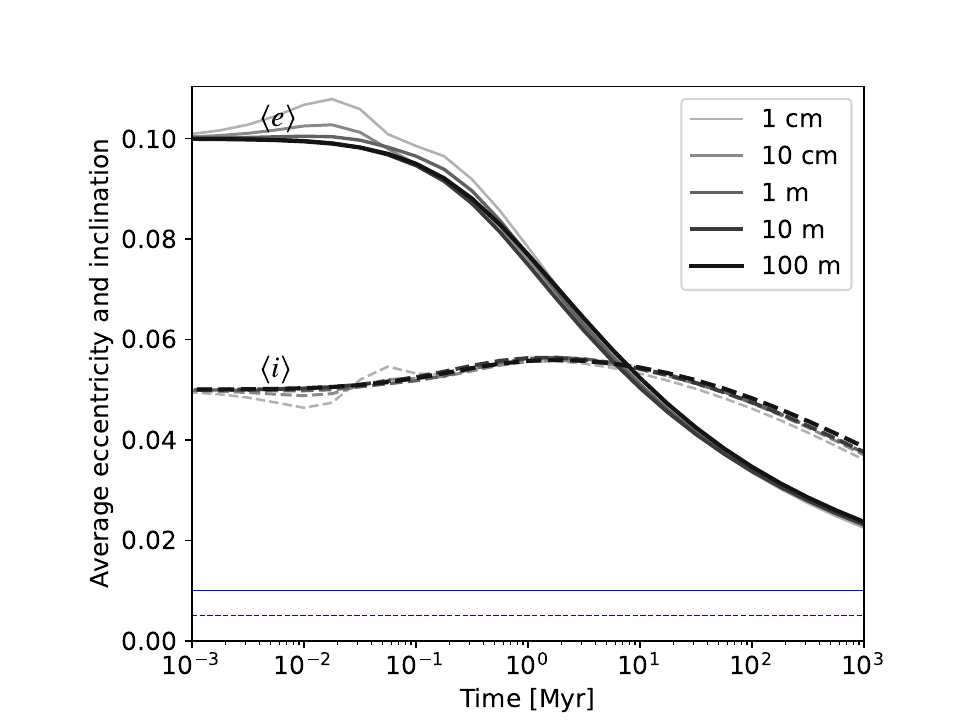}
    \caption{Same as Fig. \ref{fig:qdstar_const}, but for size-dependent material strength ($Q_{\rm D}^*$ model 2 in Table \ref{tab:qdstar}). See Sect. \ref{sec:qdstar_size_vel}.}
    \label{fig:qdstar_size}
\end{figure*}

\begin{figure*}
    \centering
    \includegraphics[width=\columnwidth]{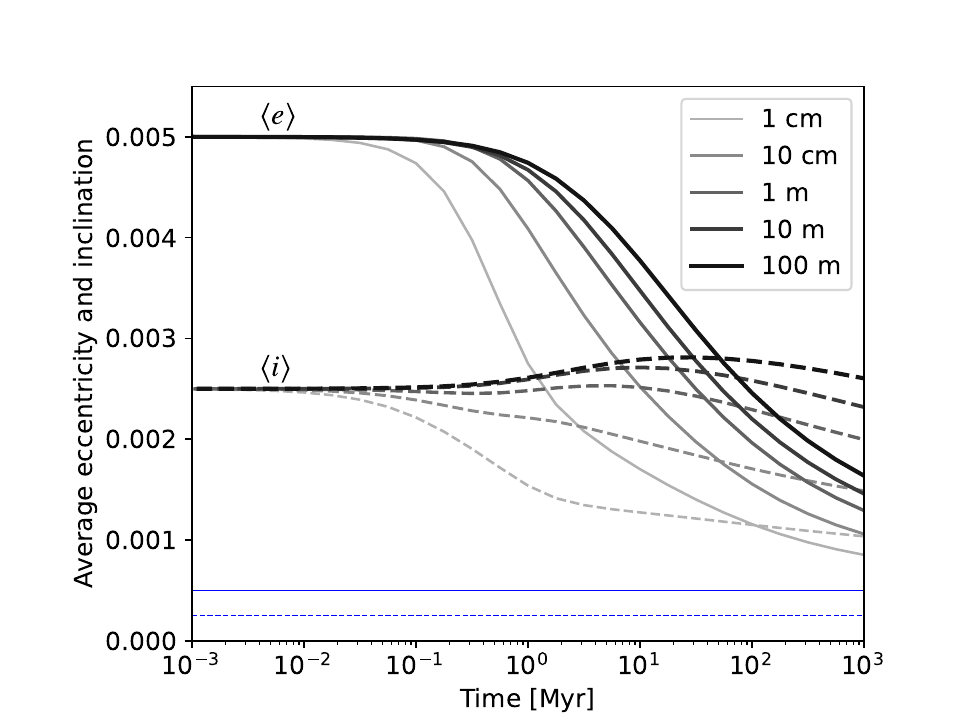}
    \includegraphics[width=\columnwidth]{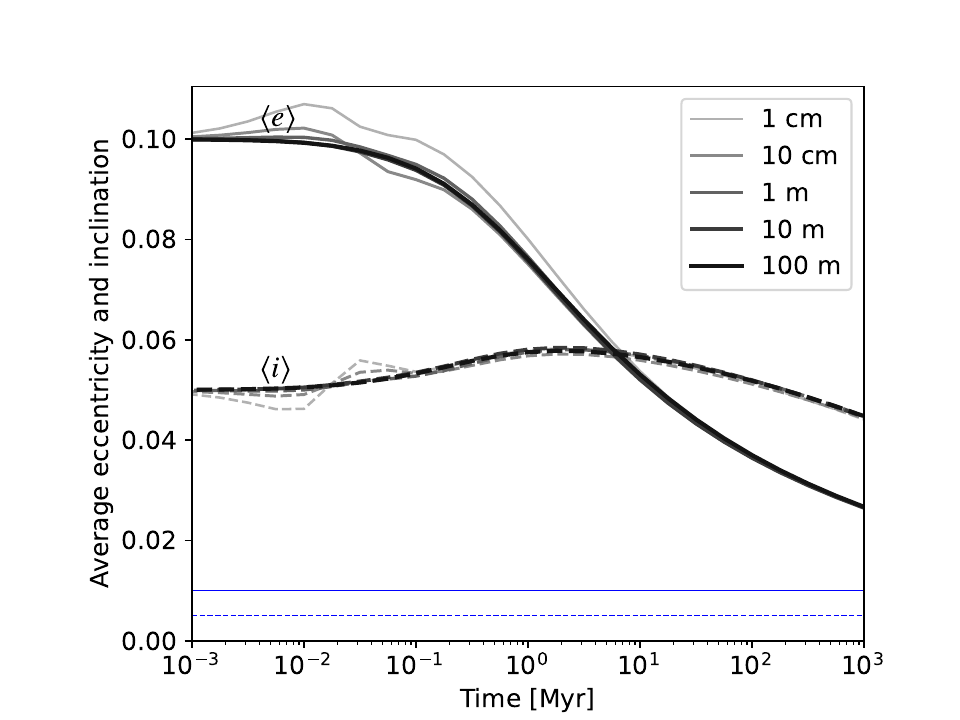}
    \caption{Same as Fig. \ref{fig:qdstar_const}, but for size- and velocity-dependent material strength ($Q_{\rm D}^*$ model 3 in Table \ref{tab:qdstar}). See Sect. \ref{sec:qdstar_size_vel}.}
    \label{fig:qdstar_size_vel}
\end{figure*}

In this section, we continue our exploration of collisional damping in collisional cascades, now with more realistic material properties. To produce the results shown in Figs. \ref{fig:qdstar_size} and \ref{fig:qdstar_size_vel}, we use models 2 and 3 for $Q_{\rm D}^*$ from Table \ref{tab:qdstar}, respectively. Model 2 for $Q_{\rm D}^*$ accounts for the size-dependence of material strength, and model 3 additionally includes the velocity dependence.

We find that the eccentricities and the inclinations in the high-excitation disc evolve similarly regardless of the exact values of $Q_{\rm D}^*$ (see the similarity in the right-hand-side panel of Fig. \ref{fig:qdstar_const} with the right-hand-side panels of Figs. \ref{fig:qdstar_size} and \ref{fig:qdstar_size_vel}). In all 3 models of the high-excitation disc the particle projectile-to-target mass ratio is much smaller than unity for all particles (see the orange lines in Fig. \ref{fig:crit_ratio}) and in all 3 models collisional damping is much less efficient than predicted by the indestructible-particles model (compare with the right-hand-side panel of Fig. \ref{fig:inbounc}).

For our low-excitation disc parameters, with these realistic prescriptions for material strength we probe an interesting regime where the expected critical projectile-to-target mass ratio is above unity for smaller particles, and below unity for larger ones (see the dashed and the dotted blue lines in Fig. \ref{fig:crit_ratio}). In such systems we expect the efficiency of damping to depend on particle size (in addition to how the collisional/damping timescale depends on particle size even if damping is efficient). Indeed, in our low-excitation disc, with our model 2 for $Q_{\rm D}^*$ (the dashed blue line in Fig. \ref{fig:crit_ratio}), we find that the velocities of the smallest bodies are damped on practically the same timescale as in our test indestructible-particles model, while for the largest bodies the average eccentricity and the inclination are not fully damped by 1\,Gyr (compare the left panel of Fig. \ref{fig:inbounc} with the left panel of Fig. \ref{fig:qdstar_size}). The differences are quite small, however, because the critical projectile-to-target mass ratio is still near unity even for the largest bodies.

In our low-excitation disc with our model 3 for $Q_{\rm D}^*$ the critical projectile-to-target mass ratio is (initially) less than one for all but the smallest particles (see the dotted blue line in Fig. \ref{fig:crit_ratio}), and in this case we find that the damping efficiency depends on particle size more strongly, albeit still weakly overall. The results for this model are shown in the left panel of Fig. \ref{fig:qdstar_size_vel}. Interestingly, for all sizes, but especially for the largest particles, the evolution of the average eccentricities and inclinations evidently resembles more the evolution in the high-excitation disc than in the previously discussed low-excitation models (compare the left and the right panel of Fig. \ref{fig:qdstar_size_vel}). In other words, there is a gradual transition in our results as a function of the projectile-to-target mass ratio across different models. In this particular model, large bodies evolve the same as in our high-excitation disc, because they are easily disrupted in collisions. Small bodies' inclinations are initially damped, but that damping slows down at one point, likely because the larger bodies are easily disrupted (on their own longer collisional timescale), producing smaller bodies with larger inclinations (because the larger bodies' inclinations are not damped).

Overall, comparing the different models presented here, we find that collisional damping becomes gradually less efficient (the damping timescale gradually increases) as the initial typical critical projectile-to-target mass ratio decreases below unity, when more bodies can be catastrophically disrupted. However, if the projectile-to-target ratio is much smaller than unity, the evolution of the average eccentricity and inclination appear to not depend on its value (see the similarity in our results for the high-excitation disc when using different $Q_{\rm D}^*$ models). In this regime any damping occurs because of how the eccentricity and inclination distributions are shaped by destructive collisions, as discussed in the previous section.

\section{Discussion} \label{sec:discussion}
In this section we discuss how our results apply to pre-stirred debris discs, implications for stirred discs, implications for small dust grains affected by radiation pressure, and lastly we discuss some of the limitations of our numerical approach.

\subsection{Pre-stirred debris discs} \label{sec:discussion_prestirred}
The results presented in this work can be directly applied to pre-stirred discs, as we assume that discs start out with some distributions of eccentricities and inclinations and we do not account for any subsequent stirring that may occur through gravitational interactions between bodies in the disc. Here we discuss what our results imply for potential observations of pre-stirred discs. While our results are not directly applicable to evolution or observations of $\sim$\,$\mu$m-sized dust grains (since such grains are not included in the model), those are discussed separately further below.

We find that if the critical projectile-to-target mass ratio for dust particles is well below unity the disc evolution is fragmentation-dominated rather than damping-dominated, the average particle inclination evolves very slowly, and it is approximately the same for all particle sizes at any given time. As observations at a given wavelength probe mostly particles of size of the order of that wavelength, these discs would be observed to have a scale-height that does not vary with observing wavelength. Therefore, a vertically thick disc with wavelength-independent scale-height does not necessarily imply a viscously stirred disc to maintain that height. 

Additionally, for such discs our model predicts that the ratio of the average eccentricity and the average inclination decreases over time to less than unity, in contrast with the value of $\langle e \rangle/\langle i \rangle \sim 2$ expected from energy equipartition in a disc of gravitationally interacting particles \citep{Ida1992}. This ratio could be observationally constrained, in principle, by measuring radial and vertical widths of narrow debris rings. If found to be unexpectedly small, it could be a sign of catastrophic collisions shaping the distributions of eccentricities and inclinations. However, it should also be recognised that such narrow rings could be different to the model discs in this work, in which we assume a non-negligible dispersion of particle semi-major axes.

If the critical projectile-to-target mass ratio approaches unity, collisional damping becomes more efficient. Because smaller particles are stronger in collisions (in the strength regime) and because their collision rate is higher, smaller particles are damped faster than larger ones. These discs may be observed to have a scale-height that increases with observing wavelength.

For the critical projectile-to-target mass ratio above unity, and for collisional timescales shorter than system age, particle inclinations are strongly damped. In observations this would appear as a very thin disc with an unresolved scale-height. The eccentricities would also be damped, but the disc would still be radially wide if its semi-major axis distribution is wide. Overall, if debris discs are in general pre-stirred, we may expect to observe two types of discs: vertically thick discs where the collision velocity is higher than the critical velocity derived above, and discs with very small vertical heights, perhaps unresolved in observations (with caveats discussed further below).

The critical projectile-to-target mass ratio is size-dependent and velocity-dependent. This has two important consequences. Firstly, in this work we only considered particles held together by material strength which become weaker with increasing size. However, above a few 100s of metres particles are held together by gravity, and they become stronger with increasing size. Therefore, at the top of the collisional cascade damping may be more efficient than for the intermediate particle sizes. This may have implications for the collisional evolution of the disc, its dust production and its observability, as discussed further below.

\begin{figure}
    \centering
    \includegraphics[width=\columnwidth]{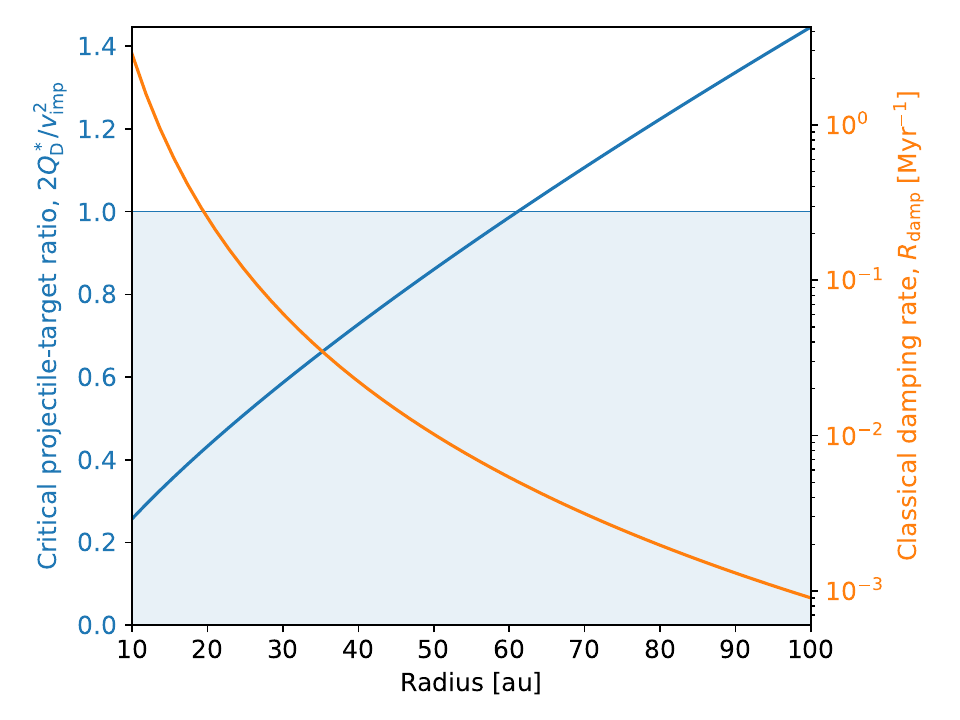}
    \caption{Example calculation of critical projectile-to-target mass ratio (in blue) and the classical damping rate given by eq. (\ref{eq:r_damp}) for 1\,mm grains as functions of radius in a wide disc around a solar-mass star, with particle $Q_{\rm D}^*$ given by model 3 in Table \ref{tab:qdstar}, an average eccentricity of $\langle e \rangle=0.01$, particles ranging from 1\,$\mu$m to 100\,m in size with a power-law size distribution exponent $q=3.5,$ and a disc mass of $0.1$\,M$_\oplus$. This example illustrates how, in a wide disc, collisional damping may be a non-monotonous function of radius: inefficient at short radii due to low projectile-to-target ratio and slow at large radii due to low collisional rate, with a `sweet spot' in the middle where damping is both efficient and fast enough to be observed assuming, for example, a system age of 100\,Myr. See Sect. \ref{sec:discussion_prestirred}.}
    \label{fig:raddep_illustration}
\end{figure}

Secondly, due to the velocity dependence of the critical projectile-to-target mass ratio, the \textit{true} damping rate (unlike the classical damping rate, $R_{\rm frag}$) may be a non-monotonous function of semi-major axis. In a wide disc it may have a minimum at some distance from the star, and so the dust scale-height may have a minimum at some distance from the star, increasing both radially inwards and outwards. This is because at shorter distances collision velocities are higher, the critical projectile-to-target mass ratio is lower and damping becomes less efficient. Conversely, if damping is efficient, the damping timescale increases radially outwards because the collision velocities and collision rate are lower radially outwards. This is illustrated with an example in Fig. \ref{fig:raddep_illustration}.

We have focused on the importance of the critical projectile-to-target mass ratio as the parameter controlling the level of collisional damping. Equivalently, given a model for $Q_{\rm D}^*(s,v_{\rm imp})$ (e.g. the commonly used expression named model 3 in Table \ref{tab:qdstar}), damping will be inefficient if collision velocities are above a critical impact velocity that is a solution to $1 = 2 Q_{\rm D}^*(s,v_{\rm imp})/v_{\rm imp}^2$. For mm-sized dust grains, using the model 3 from Table \ref{tab:qdstar}, we find a critical velocity of $v_{\rm crit} \sim$\,40\,m\,s$^{-1}$. We use this value in the discussion further below, but note that it is derived using a $Q_{\rm D}^*$ model designed for much higher ($\sim$\,km\,s$^{-1}$) impact velocities. Therefore, this specific threshold could change if a collisional model more suitable for low-velocity disruptive impacts were adapted.

Using this critical velocity and the simple collisional model discussed in Sect. \ref{sec:theory}, we can consider if collisional damping is efficient or inefficient (relative to the classical damping rate given by eq. (\ref{eq:r_damp})) in different types of debris discs. The typical collision velocity in a disc can be estimated as $v_{\rm rel} \sim v_{\rm kep} i \sim v_{\rm kep} h$, where $i$ is the typical inclination and $h$ the aspect ratio of the disc (and the latter can be measured for some discs). Thus, for a given mass of the central object $M_*$ and disc radius $a$, there is a critical value of the aspect ratio, $h_{\rm crit} = v_{\rm crit}/v_{\rm kep}$. For an exo-Kuiper belt ($M_*=2$\,M$_\odot$, $a=100$\,au), $h_{\rm crit}=0.01$; for a disc around a white dwarf ($M_*=0.6$\,M$_\odot$, $a=1$\,R$_\odot$), $h_{\rm crit} = 10^{-4}$; for a planetary ring ($M_*=1$\,M$_{\rm Saturn}$, $a=10^5$\,km), $h_{\rm crit}=0.002$. If $h<h_{\rm crit}$, collision velocities are small in the system and collisional damping is efficient. This does not necessarily mean that such a disc would be observed to be very thin and unresolved, because that would still require the damping timescale to be shorter than the system age. In our example exo-Kuiper belt (assuming disc mass of $M=1$\,M$_\oplus$, $\Delta a/a=0.2$, $\rho_{\rm b}=3$\,g\,cm$^{-3}$, $s_{\rm max}=1$\,km), the damping timescale of mm-sized dust grains is about 30\,Myr, so damping may be ongoing in young exo-Kuiper belts.

If $h>h_{\rm crit}$, collision velocities are high and damping is inefficient. However, this does not mean that the typical inclination or the disc thickness has not decreased within the system age. Even in high-excitation discs we found eccentricities and inclinations can decrease due to preferential destruction of particles on certain orbits. This decrease occurs on a timescale much longer than the classical damping timescale of the largest bodies in the disc. For exo-Kuiper belts this \textit{true} damping timescale is too long, but for other debris systems that timescale may still be much shorter than the system age. In our example discs around a white dwarf and a planetary ring, assuming a disc mass of $M=1$\,M$_{\rm Moon}=0.01$\,M$_\oplus$ (and other disc parameters as above), the classical damping rate of 1\,km bodies evaluates to $10^{-3}$ and $10^{-4}$\,yr, respectively. These are tiny fractions of Gyr-long lifetimes of these debris discs and so, even if it takes much longer to reduce the average particle inclination, we may expect that the disc thickness has decreased significantly within the system lifetime, despite the high collision velocities.

Furthermore, to be observable, debris discs also need to remain bright enough, that is, the fragmentation timescale of the largest bodies in the disc must not bee too short, or else the disc loses mass too quickly \citep[e.g.][]{Wyatt2007a}. On the other hand, if a disc initially consists only of large bodies, their fragmentation timescale also needs to be short enough for the bodies to have produced the observable small dust; in this case there is a lower limit on the initial excitation of the disc. Furthermore, to observe any damping, the damping timescale (the real one, not the idealised one from eq. (\ref{eq:r_damp})) must be shorter than the system age. However, if the damping timescale is too short and particle velocities are severely damped, collisions of large bodies would not produce the small dust that can be observed. Growth, instead of fragmentation, may then govern the evolution of the disc, reducing the disc luminosity \citep[][]{Kenyon2002a}, followed potentially by a revival if this leads to formation of (dwarf) planets that can stir the disc \citep{Kenyon2004a}. Even before damping occurs in such a disc there may be very little dust - since efficient damping requires a critical projectile-to-target mass ratio of unity or higher, such a disc would rely on cratering to create smaller particles. These timescales depend on several poorly constrained factors, such as the size of the largest bodies present in the disc, see eq. (\ref{eq:r_coll_constant}). 

Several previous studies considered the evolution of mass and velocity in pre-stirred discs. \citet{Kenyon2002a} and \citet{Kenyon2004a} found that very-high-excitation discs lose mass and become non-observable before inclinations are damped. This is consistent with the fragmentation rate being much higher than the damping rate if the critical projectile-to-target mass ratio is much less than unity. In less extreme cases, they found that making bodies weaker led to particles damping more rapidly. This appears to be a consequence of assuming that damping acts 
at the rate equivalent to the one given by eq. (\ref{eq:r_damp}) in catastrophic collisions. However, if mass and velocity evolution are fully coupled and both catastrophic and cratering collisions are assumed to be fully dissipating, as in our simulations, the opposite occurs (making bodies weaker results in slower damping). It is less clear why in their results the eccentricity is found to decrease with increasing particle size for the smallest particles. This may be due to the collisional/damping timescale for the smallest particles being increased due to the lower-size cutoff (we observe this effect to some degree in our simulations), or perhaps due to other effects included in their simulations and not in ours, such as dynamical friction.

Furthermore, \citet{Krivov2005} studied pre-stirred discs using the same sort of kinetic approach as in our work, except that they did not include orbital inclinations in their phase space. Their disc models are similar to our high-excitation, fragmentation-dominated discs in which the eccentricity distribution is shaped by how the collisional probability depends on the eccentricity. They found that particles on orbits with both the smallest and largest eccentricities are preferentially depleted, in contrast to our models where only particles with the largest eccentricities are preferentially removed. This is because the collision probability in their models, calculated analytically, increases for low eccentricities. However, in a belt that is wider than the typical eccentricity, we do not expect and do not observe that dependence (see Fig. \ref{fig:qdstar_const_rate}).

\subsection{Viscously stirred discs}
Our numerical simulations model pre-stirred discs, but  many studies have argued that the large bodies present in debris discs gravitationally stir all particles in the disc, and that this viscous self-stirring can explain the levels of excitation in the observed debris discs \citep[e.g.][]{Kenyon2002b, Kenyon2004a, Kenyon2008, Kenyon2010, Kennedy2010, Krivov2018b}. Damping mechanisms can still act in viscously stirred discs, although it is often found in numerical simulations that they are not significant if very large bodies are formed through accretion \citep{Kenyon2002b, Kenyon2004a, Kenyon2008, Kenyon2010}. For such discs our results imply that collisional damping becomes less and less important over time as the velocities in the disc become more excited by viscous stirring.

\citet{Pan2012} argued, using an analytical model, that the velocity dispersion (or, equivalently, the inclination dispersion) of particles of different sizes is determined by an equilibrium between viscous stirring and collisional damping. For bodies participating in the collisional cascade (all except for, maybe, the largest bodies) and for $q<4$ ($q$ being the size distribution exponent), they considered two types of damping: collisional damping through collisions with equal-sized bodies and collisional damping through catastrophic collisions. Based on our results, the former would occur only if the projectile-to-target size ratio is of the order unity or larger. Notably, in that regime collisional cascades might be driven purely by cratering/erosion, which might require a different form of mass conservation to be considered, compared to the one used in their study.

If the projectile-to-target size ratio is less than one (or it becomes less than one due to stirring), the second type of damping applies. However, it should be noted that velocities are not really damped in this case. Rather, it should be understood as the case in which particles are destroyed at the same rate at which they are stirred. That is analogous to our study of the balance between damping and destruction rates. However, it is unlikely that stirring can be balanced in this way. Indeed, it could occur that particles of a certain size are being destroyed faster than their velocities can be increased by stirring, but at the same time these particles would be continually produced in collisions of larger bodies. Newly created fragments would have a velocity dispersion inherited from their parent bodies, which is not accounted for by the analytical model. If the critical projectile-to-target mass ratio is much less than one, fragments in the collisional cascade inherit the velocities of the large bodies. Since these large bodies can be stirred to higher velocities (their damping and fragmentation rates are lower), the small particles would also be effectively stirred to higher velocities than implied by the equilibrium calculated using this analytical model. Ultimately, the velocities of all particles may be equal to the velocities of particles at the top of the collisional cascade in this case.

Either way, it should also be noted that the equilibrium might not be reached in the observed debris discs. For example, the fragmentation rate might be lower than the viscous rate, but fast enough to deplete the disc before it is stirred to the equilibrium value. The viscous stirring rate may also increase over time as the largest bodies grow \citep{Kenyon2002b, Kenyon2004a, Kenyon2008, Kenyon2010}.

\subsection{Small dust grains}
The smallest sub-micron-sized dust grains produced in collisions in debris discs are expelled from the disc by radiation pressure, on orbital timescales, and the smallest micron-sized grains that remain bound are put on high-eccentricity orbits \citep[e.g.][]{Krivov2006}. As a result, their fragmentation timescale is fairly independent of the level of excitation in the disc \citep{Thebault2008}. Due to these high eccentricities, the critical projectile-to-target size ratio is likely to be less than unity for these particles \citep[although material strength is found to increase with decreasing particle size for small bodies, and these micron-sized grains are inferred to be be quite strong;][]{Benz1999}. Therefore, based on our results in this paper, we hypothesise that collisional damping is inefficient for these particles. Instead, these particles likely inherit their inclinations from their parent bodies. Their observed scale-height may decrease with time if the inclinations of their parent bodies are damped. That would occur on the fragmentation timescale of the parent bodies.

It has been proposed that if physical collisions of these particles are not maximally damping (i.e., if there is some elasticity), some of the kinetic energy associated with their high eccentricities is redistributed into vertical direction, so that collisions increase their inclinations instead of damping them \citep{Thebault2009}. A consequence of this would be, for example, that the average inclination decreases as a function of size for these particles, which could be measured with scattered light observations. Implications of our findings are not straight-forward in this case. On the one hand, numerical models of \citep{Thebault2009} only included bouncing collisions and we expect these particles to fragment and become smaller in collisions, instead of bouncing off and increasing their inclination dispersion. Any increase in the scale-height should be reflected in the distribution of their fragments, which are likely to be unbound and leave the system quickly. On the other hand, it has been argued by \citet{Thebault2009} that these dust grains may act as projectiles in collisions in which grains of the same size are ultimately produced as fragments, with similar inclinations as would result from bouncing collisions. This deserves further study with a numerical method that incorporates both the velocity and mass evolution of these particles - such as the method employed in our work.


\subsection{Limitations} \label{sec:limitations}
Throughout the paper we stressed the advantages of the kinetic approach \citep{Krivov2005,Krivov2006,vanLieshout2014} used in this work. It allows to model concurrently and in a fully coupled manner both the mass and the velocity evolution of a circumstellar disc. The results presented show the importance of this in our understanding of the collisional evolution of debris discs. However, there are several important limitations.

The kinetic approach considers the evolution of particles in a phase space of particle size and select orbital elements. For an axisymmetric disc it is sufficient to consider semi-major axis, eccentricity and inclination. The numerical approach necessitates discretisation of this phase space. As a consequence, changes in particle eccentricities and inclinations that are smaller than the bin widths are neglected. Additionally, if a particle moves to a different bin, it is instantly assumed to be moved to the centre of that bin. To check how this influences our results, we run some of our models at different grid resolutions, that is, with different number of eccentricity and inclination bins. These tests are shown and discussed in Appendix \ref{sec:tests_km}. In short, we find that for those models where damping is efficient (critical projectile-to-target ratio of order unity or larger) the damping rate in our models is to some degree influenced by grid resolution: with smaller bin widths, damping is slightly faster. In spite of that, our main conclusions still hold, and the lack of significant damping when the critical projectile-to-target ratio is much smaller than one does not appear to depend on grid resolution.

Furthermore, in this work we neglected the evolution of semi-major axis. This has been justified by the findings of \citet{Brahic1977} and \citet{Lithwick2007} that semi-major axes evolve on much longer timescales than eccentricities and inclinations, and this study focuses on the evolution of the latter. However, for real debris discs, it is still possible that the semi-major axis distribution may evolve within system ages. This is especially true for smaller bodies whose collisional timescales are short. General expectations for that case have been outlined by \citet{Brahic1977}, who show how long-term spreading of the disc eventually leads to some particles falling onto the star, as well as to a decrease of collisional rates. In the case of high-excitation discs in which collisional damping is inefficient, the findings of \citet{Krivov2005} apply: particles at shorter semi-major axis are preferentially depleted due to shorter collisional timescales. 

There are several other assumptions in our work that may need to be questioned in future work. We assume that particles have initially uniform distributions of eccentricities and inclinations, whereas assuming something akin to a Rayleigh distribution may be more appropriate \citep[e.g.][]{Brahic1977,Ida1992,Lissauer1993}. Ideally, after some time these distributions are independent of the initial conditions in the model, but this might also not be the case in the models where they evolve slowly, such as our high-excitation disc models. Furthermore, we neglect any elasticity in collisions, assuming maximal efficiency of collisional damping in each single collision. The impact energy is assumed to be fully dissipated and the velocity of all particles post-collision is assumed to equal the pre-collision centre-of-mass velocity of the colliders; orbits of both remnants and fragments in both catastrophic and cratering collisions are determined by this. This assumption maximises the rate at which the average eccentricity and inclination decrease with time. In reality, the level of dissipation likely depends on the collisional outcome and collision geometry \citep[e.g. catastrophic collisions dissipate more energy than cratering ones, and more energy is dissipated in head-on versus grazing impacts;][]{Fujiwara1986,Leinhardt2012}. We expect that our results in the high-velocity regime would not change with different dissipation assumptions, because in that regime the low projectile-to-target mass ratio makes damping insignificant in any case. At low collision velocities damping could be less significant in real discs than it is in the results of our simulations, although it would always be present to some degree as collisions are always inelastic.
All in all, most of these limitations can be addressed in future work, and the general kinetic approach used in this work is a very important tool for studying the collisional mass and velocity evolution of debris discs.

\section{Conclusions} \label{sec:conclusions}
We present a kinetic model of the collisional evolution of debris discs that includes the evolution of orbital inclinations in addition to the evolution of particle masses and orbital eccentricities. We applied this model to pre-stirred discs with different levels of excitation and with particles of different material strengths. We explored how inelastic collisions change particle eccentricities and inclinations over time. Our main results can be summarised as follows:
\begin{enumerate}
  \item Our model reproduces the basic theoretical expectations regarding the collisional damping of indestructible particles  in discs. In discs in which particle mass can be changed by cratering and catastrophic disruption, the critical projectile-to-target mass ratio determines how efficient collisional damping is.
  \item If the critical projectile-to-target mass ratio is of order unity or larger (low-excitation discs and/or particles with high material strength), collisions mostly leave colliders intact and particle eccentricities and inclinations are damped over time at a similar rate to the case where particles are indestructible.
  \item As the critical projectile-to-target mass ratio falls below unity, the damping timescale becomes longer. This is because collisional damping is most efficient in collisions of bodies of similar mass, and in this case those collisions result in catastrophic disruption. Collisions with projectiles smaller than the critical size leave targets largely intact, but they do not result in significant damping, and overall the destruction rate is higher than the damping rate.
  \item If the critical projectile-to-target mass ratio is much smaller than one (high-excitation discs and/or discs in which particles are made of weak materials), eccentricities and inclinations evolve slowly, and their evolution is driven by the destruction probability being slightly different for particles on different orbits. In this regime, the damping rate does not depend on the exact value of the projectile-to-target ratio in the range explored. Furthermore, the average eccentricity and inclination are independent of size at all times, meaning that the scale-height of such a disc would appear the same at different wavelengths. We also find that the average eccentricity decreases faster than the average inclination.
  \item For mm-sized dust grains, we expect the critical projectile-to-target mass ratio to fall below unity at impact velocities higher than $\sim$\,40\,m\,s$^{-1}$. For impact velocities much higher than this critical velocity, we expect collisional damping to be inefficient in changing the velocity distribution with which mm-sized grains are produced, or in counteracting viscous stirring from large bodies in the disc. 
  \item Both the collisional timescales and the critical projectile-to-target mass ratio are important factors in determining whether collisional damping can significantly affect eccentricities and inclinations and vertical scale-heights in a debris disc.
\end{enumerate}

\begin{acknowledgements}
We thank the referee for their helpful and constructive comments. We thank Jessica Rigley, Rik van Lieshout, Sebastian Marino and Nicole Pawellek for helpful discussions. MRJ acknowledges support from the European Union's Horizon Europe Programme under the Marie Sklodowska-Curie grant agreement no. 101064124, funding provided by the Institute of Physics Belgrade, through the grant by the Ministry of Science, Technological Development, and Innovations of the Republic of Serbia, and support by the UK Science and Technology Research Council (STFC) via the consolidated grant ST/S000623/1.
\end{acknowledgements}

\bibliographystyle{aa} 
\bibliography{bibliography} 



\begin{appendix}

\section{Tests of the Monte Carlo code} \label{sec:tests_mc}
\begin{figure}
        \includegraphics[width=\columnwidth]{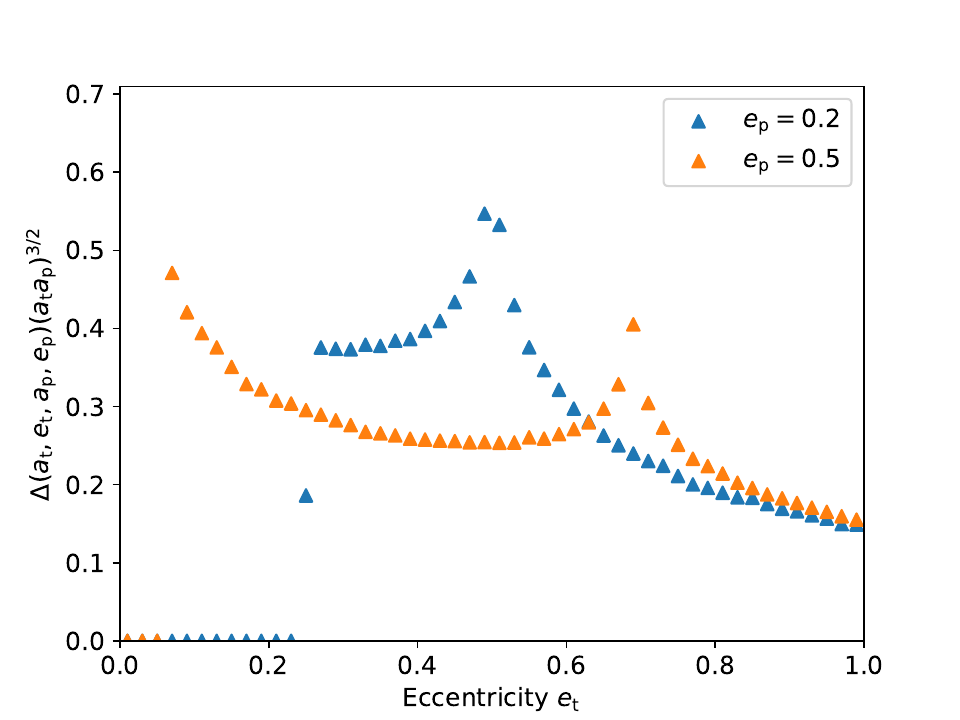}
        \includegraphics[width=\columnwidth]{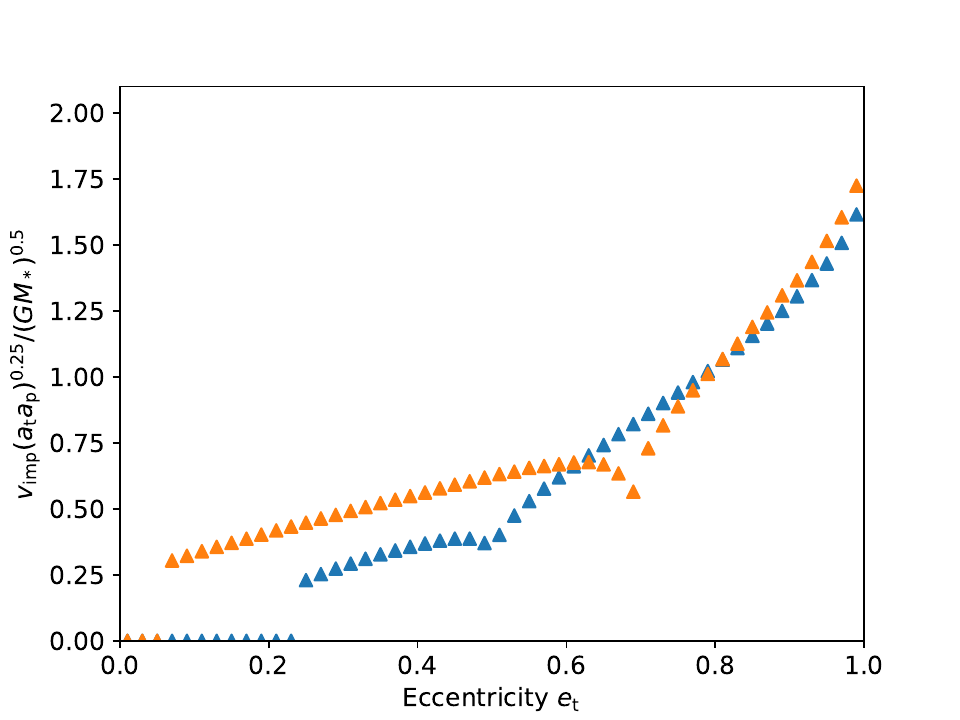}
    \caption{Test of the Monte Carlo code in 2D: geometric probability $\Delta$ (top panel) and the average impact velocity $\bar{v}_{\rm imp}$ (bottom panel), calculated for test populations with semi-major axes $a_{\rm t} = 1.6, a_{\rm p} = 1.0$ and assumed semi-opening angle $\epsilon=8.5^{\circ}$, using the Monte Carlo code in two dimensions, for comparison with figure A.3 of \citet{Krivov2006}.}
    \label{fig:test_krivov}
\end{figure}

We first test the Monte Carlo calculation of collision probabilities and impact velocities in two dimensions, by reproducing Fig. A.3 of \citet{Krivov2006}, see Fig. \ref{fig:test_krivov}. The top panel shows the $\Delta$-integral defined by \citet{Krivov2006}, and the bottom panel shows the average impact velocity $\bar{v}_{\rm imp}$ between two test populations of particles on planar, eccentric orbits, calculated as (see eq. (31), (32) in \citet{Wyatt2010} and Sect. \ref{sec:methods_mc})
\begin{align}
  \Delta &= \sum_{r,\phi} \sigma_1(r,\phi) \sigma_2(r,\phi) / dV(r,\phi), \\
  \bar{v}_{\rm imp} &= \sum_{r,\phi} \sigma_1(r,\phi) \sigma_2(r,\phi) \langle v_{\rm imp}(r,\phi)\rangle / dV(r,\phi) / \Delta.
\end{align}
To obtain their figure A.3, \citet{Krivov2006} used 3D Monte Carlo simulations in which they assumed uniform distributions of particle latitudes. To obtain our Fig. \ref{fig:test_krivov}, we used 2D Monte Carlo simulations which implicitly assume the same, uniform distributions of particle latitudes. Our results do not account for the vertical components of particle velocities, but these are negligible for these test parameters, and the agreement between their results and ours is very good.

\begin{figure}
        \includegraphics[width=\columnwidth]{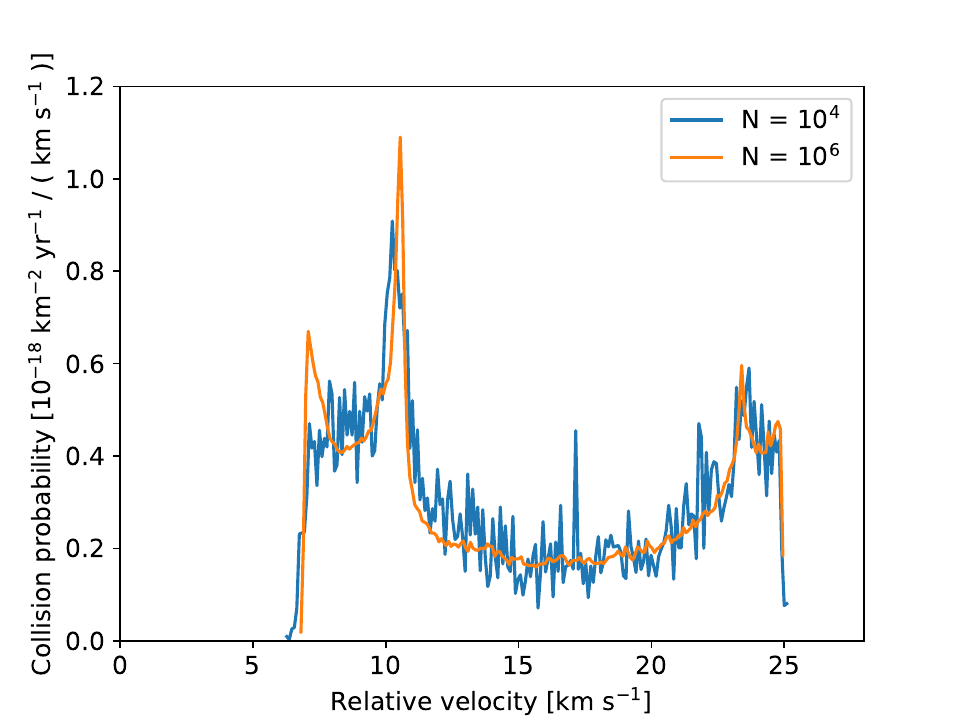}
    \caption{Test of the Monte Carlo code in 3D: distribution of intrinsic probability as a function of impact velocity between two test populations of particles with orbital elements $a_{\rm t} = 3.42$\,au, $e_{\rm t} = 0.578$, $i_{\rm t} = 0.435$\,rad, and $a_{\rm p} = 1.59$\,au, $e_{\rm p} = 0.056$, $i_{\rm p} = 0.466$\,rad, obtained using the Monte Carlo code, using $10^4$ and using $10^6$ particles in each population for the blue and the orange line, respectively, for comparison with figure 1 of \citet{Bottke1994}, see also figure 4 of \citet{Wyatt2010}.}
    \label{fig:test_bottke}
\end{figure}

Next, we test the Monte Carlo code in three dimensions by reproducing (with reasonably good accuracy) a distribution of relative velocities between two populations of particles on inclined orbits given in figure 1 of Bottke et al. 1994, see Fig. \ref{fig:test_bottke}. This figure has already been reproduced by \citet{Wyatt2010}, who first published the Monte Carlo code used here. Here the figure is recalculated directly from the output that feeds as input to the kinetic model, thereby testing the connection between the codes, and we also use it to illustrate the effect of varying the number of particles in the simulation. Throughout this paper we use $N=10^4$ particles in each population in each Monte Carlo simulation, and Fig. \ref{fig:test_bottke} shows that such sparse sampling is sufficiently accurate to reasonably approximate the velocity distribution and collision probability of colliding particles. 

\section{Tests of the kinetic model} \label{sec:tests_km}
\begin{figure}
    \centering
    \includegraphics[width=\columnwidth]{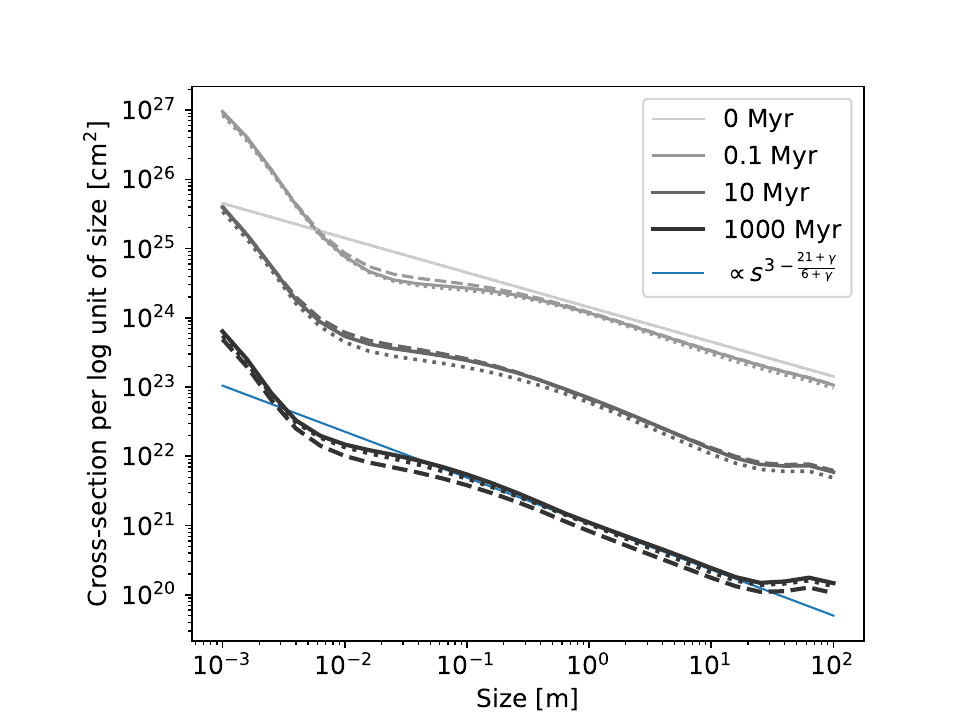}
    \caption{Cross-section area per base-10 logarithmic unit of size  for `2D' models, for our high-excitation disc parameters, obtained using Monte Carlo collision rates (solid lines) and using analytically calculated collision rates (dashed lines), and also for a corresponding `3D' model (dotted lines). Results from the three models are shown as function of time (various shades of grey, see plot legend). Also shown is an analytically derived, arbitrarily scaled power-law describing a disc at steady-state, with bodies' critical specific disruption energy in the strength regime ($\gamma=-0.37$; thin blue line).}
    \label{fig:comp_ana_tab}
\end{figure}

\begin{figure}
    \centering
    \includegraphics[width=\columnwidth]{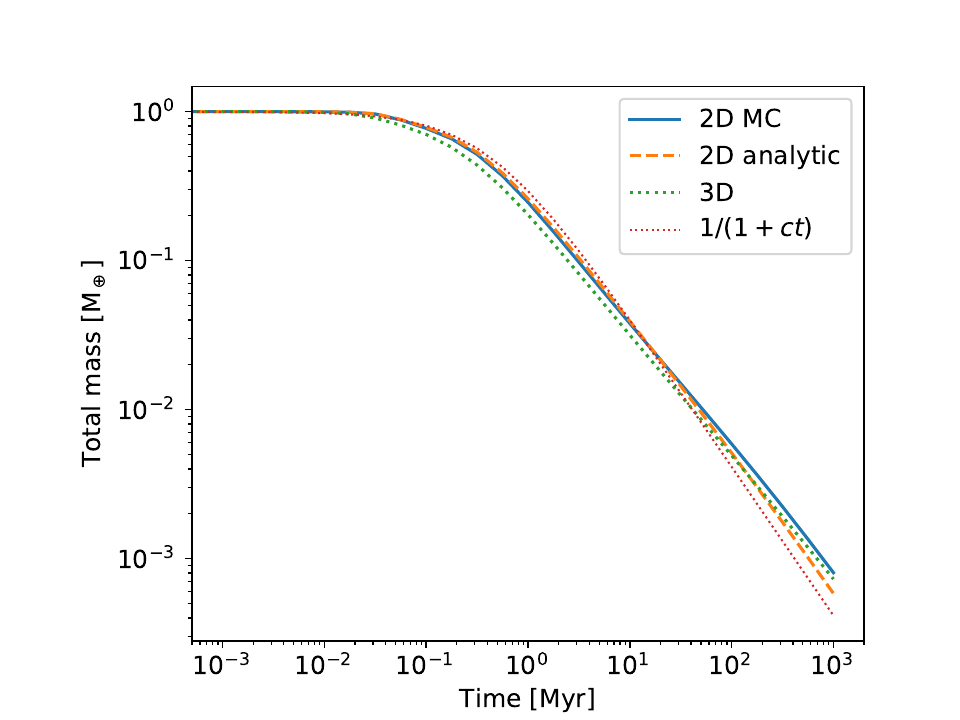}
    \caption{Total mass in the disc as function of time for the `2D' and the `3D' models, for our high-excitation disc parameters, and also the basic expectation for the evolution of disc mass from an analytic model \citep{Wyatt2007a}, see plot legend.}
    \label{fig:comp_ana_tab_mass}
\end{figure}

Under the assumptions that a disc is vertically thin and that the vertical distribution of particles is fixed and uniform in particle latitude, collision rates can be calculated in `2D' using analytic formulae with approximate corrections for the vertical component \citep{Krivov2006}. To model the evolution of the disc, it is then sufficient to consider the phase space consisting only of semi-major axis (or pericentre distance) and eccentricity. In Figs. \ref{fig:comp_ana_tab} and \ref{fig:comp_ana_tab_mass} we compare results of such `2D' models, obtained using the Monte Carlo approach presented in this paper and using the analytic approach as implemented by \citet{vanLieshout2014} with minor corrections. The parameters of the models are the same ones used throughout the paper for our high-excitation disc, and for the realistic specific critical disruption energy given by the model 3 in Table \ref{tab:qdstar}. We find that the results agree well.

Also shown in Figs. \ref{fig:comp_ana_tab} and \ref{fig:comp_ana_tab_mass} are results of our full `3D' model of a high-excitation disc. In comparison with `2D' models, this model may be expected to feature somewhat faster evolution, because the `3D' model assumes a uniform distribution of inclinations, meaning the distribution of particle latitudes is strongly peaked around the disc midplane, leading to higher collision probabilities \citep{Wyatt2010}. Nevertheless, the distribution of inclinations in this particular `3D' model remains fairly flat throughout the duration of the simulation, and the results agree well with `2D' models.

We further compare all of these results with two analytic results on the evolution of collisional cascades. Firstly, in Fig. \ref{fig:comp_ana_tab}, the thin blue line shows the power-law size distribution expected in a steady-state disc in which the specific critical disruption energy depends on the particle size as a power law, $Q_{\rm D}^*(s) \propto s^{\gamma}$, and in which the particle velocities are independent of particle size \citep[see e.g.][we insert $\gamma=-0.37$, corresponding to the strength regime]{Pan2012}. The assumption of the particle velocities being independent of particle size is appropriate because the eccentricities and the inclinations do not evolve significantly in these high-excitation disc models. We find that over time all three models evolve to this power-law size distribution, with the exception of the smallest particle sizes. The relative uptick in the number of smallest-size particles is a well known feature of collisional cascades with a lower-size cutoff \citet[e.g.][]{CampoBagatin1994,Krivov2006,Thebault2007}. In real debris discs such a feature is expected at $\sim \mu$m sizes, where the size distribution is expected to be cut off by radiation pressure. In our models the size cutoff and this feature are artificial numerical necessities. 

Secondly, it is well known that the total mass of a debris disc should decrease as a function of time approximately as $M(t)=M_0/(1+c t)$ \citep{Dominik2003,Wyatt2007a,Lohne2008}. Here the constant $c$ has the meaning of the inverse of the initial collisional timescale of the largest bodies in the disc. In Fig. \ref{fig:comp_ana_tab_mass}, we compare the mass evolution of discs from the three `2D' and the `3D' models with this functional form, shown with the thin red dotted curve. For this we use a value of $c$ obtained by fitting the thick green dotted curve (the `3D' model). Therefore, the overall shape of $M(t)$ from our models is in reasonably good agreement with the functional form derived analytically.


\begin{figure}
    \centering
    \includegraphics[width=\columnwidth]{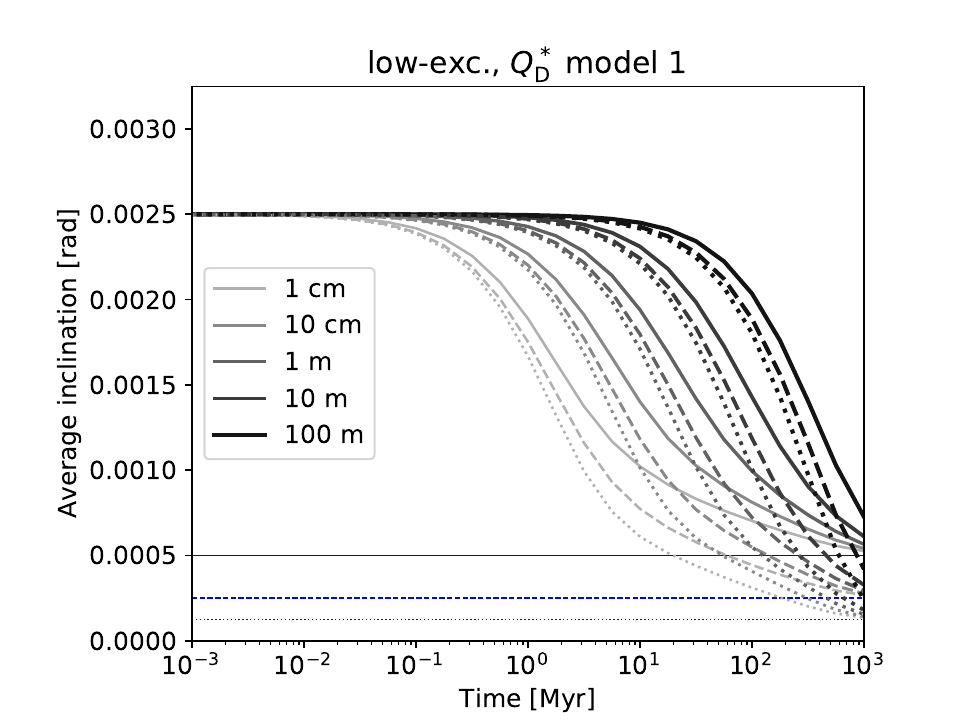}
    \includegraphics[width=\columnwidth]{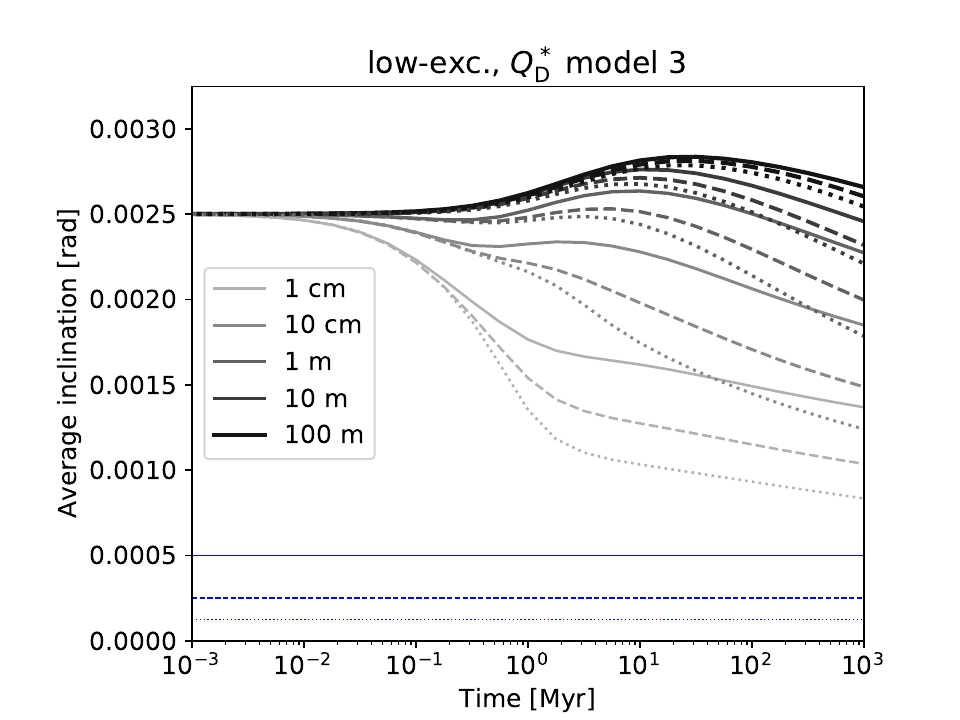}
    \includegraphics[width=\columnwidth]{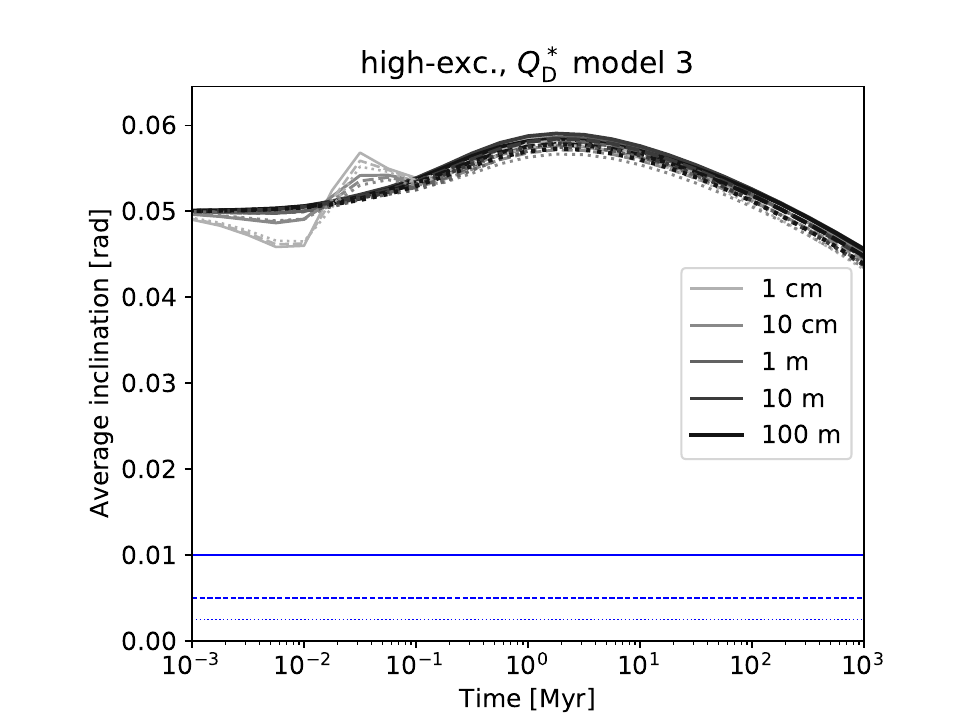}
    \caption{Resolution tests show average inclination as function of time, for different particle sizes (see plot legend), for three of our models (top panel: low-excitation disc using $Q_{\rm D}^*$ model 1; middle panel: low-excitation disc using $Q_{\rm D}^*$ model 3; bottom panel: high-excitation disc using $Q_{\rm D}^*$ model 3) for three different grid resolutions (solid lines: $n_e=n_i=5$; dashed lines: $n_e=n_i=10$; dotted lines: $n_e=n_i=20$). Horizontal blue lines show the lowest grid bin values, i.e. the lowest values to which the average inclination can be damped in each model.
    }
    \label{fig:test_res}
\end{figure}

Finally, we test how our results depend on the grid resolution. In Fig. \ref{fig:test_res} we show results for the evolution of average inclinations for 3 of our models presented in this paper with varying number of bins in the eccentricity ($n_e$) and the inclination ($n_i$) grids. We pick these 3 models so that we test convergence in 3 different evolution scenarios we encounter: efficient damping (top panel in Fig. \ref{fig:test_res}, shown in the main text in left panel in Fig. \ref{fig:qdstar_const}), inefficient damping (bottom panel in Fig. \ref{fig:test_res}, and right panel in Fig. \ref{fig:qdstar_size_vel}) and a regime in between (middle panel in Fig. \ref{fig:test_res}, and left panel in Fig. \ref{fig:qdstar_size_vel}).

We find that some of our results (in the top and the middle panel of Fig. \ref{fig:test_res}) do vary somewhat with the grid resolution. This is because of the limitation of the kinetic approach where we only consider particles to move to a different bin if the level of damping in a single collision is sufficiently large. Therefore, higher resolution (smaller bins) results in faster damping in our models, and the damping timescales are overestimated in our models. Nevertheless, there appears to exist a trend of convergence with increasing number of eccentricity and inclination bins, and the difference between $n_e=n_i=10$ results (the resolution used throughout the paper) and $n_e=n_i=20$ results is not too large with respect to all of the other uncertainties present in these numerical models. Importantly, the qualitative picture we established in the main text does not change with grid resolution: efficiency of collisional damping decreases with increasing collision velocities and decreasing material strength (i.e. there is a clear change in damping efficiency going from the top to the bottom panel of Fig. \ref{fig:test_res}, at any of the tested grid resolutions).

\section{Flowchart of numerical model}
\begin{figure}
    \centering
        \includegraphics[width=0.9\columnwidth]{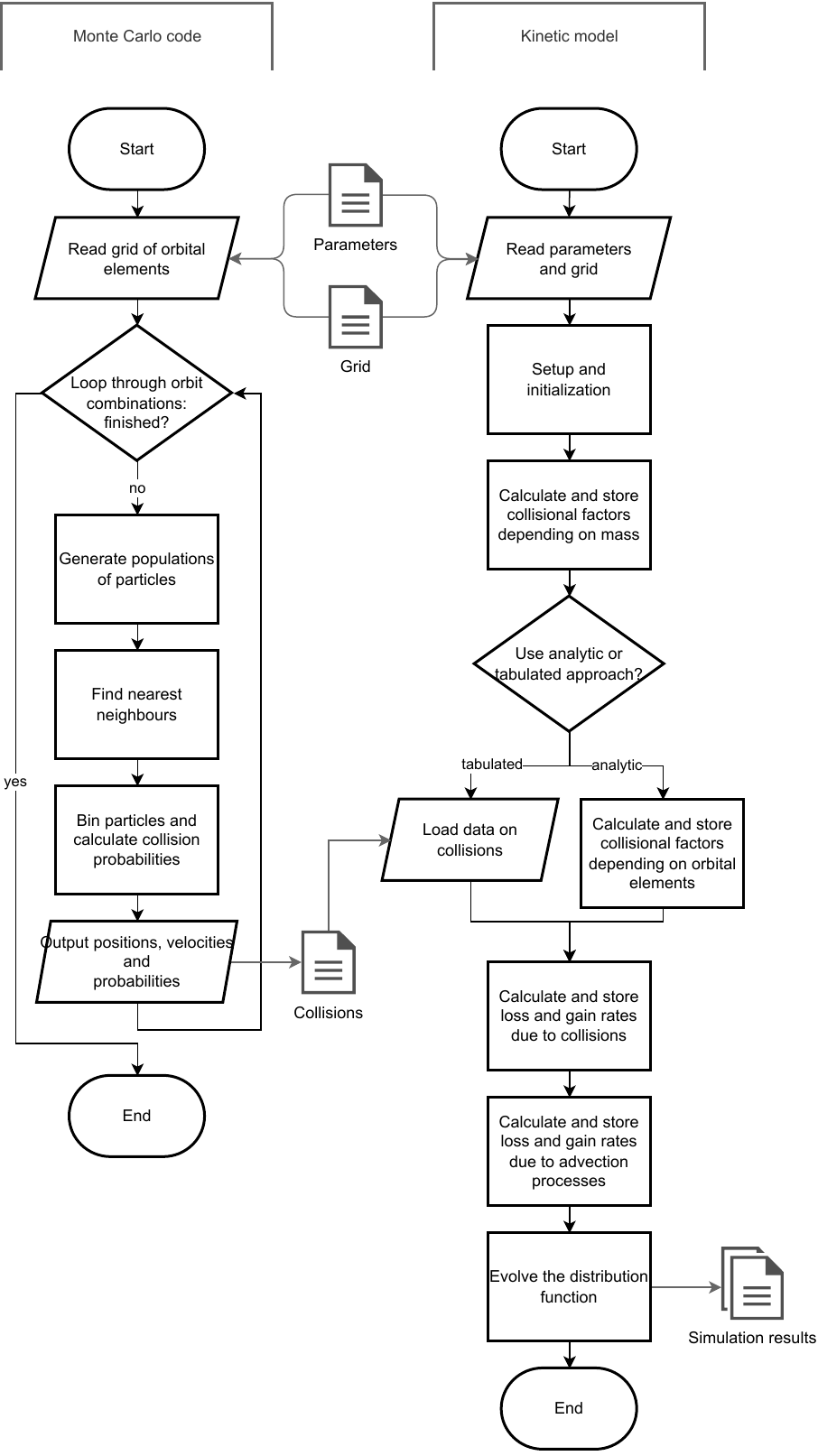}
    \caption{Flowchart illustrating the numerical model used in this work: Monte Carlo simulations \citep[][left column]{Wyatt2010} are used to pre-calculate samples of collisional pairs and collision probabilities, and these are fed as input to the kinetic model of collisional evolution \citep[][right column]{vanLieshout2014}.}
    \label{fig:flowchart}
\end{figure}

Figure \ref{fig:flowchart} illustrates the details of the numerical calculations in our model, including how the Monte Carlo simulations of \citet{Wyatt2010} and the kinetic model of \citet{vanLieshout2014} are coupled.

\end{appendix}

\end{document}